\documentclass[a4paper,12pt, epsfig]{article}
\pdfoutput=1
\usepackage{epsfig}
\usepackage{amssymb,latexsym}
\usepackage{amsfonts}
\usepackage{amsmath}
\usepackage[colorlinks, linkcolor=blue, citecolor=blue, urlcolor=blue]{hyperref}
\usepackage{epstopdf}
\usepackage{graphicx}
\newskip\humongous \humongous=0pt plus 1000pt minus 1000pt

\newif\ifdtup

\catcode`\@=11

\@addtoreset{equation}{section}
\def\theequation{\arabic{section}.\arabic{equation}}

\def\@normalsize{\@setsize\normalsize{15pt}\xiipt\@xiipt
\abovedisplayskip 14pt plus3pt minus3pt%
\belowdisplayskip \abovedisplayskip
\abovedisplayshortskip \z@ plus3pt%
\belowdisplayshortskip 7pt plus3.5pt minus0pt}

\def\small{\@setsize\small{13.6pt}\xipt\@xipt
\abovedisplayskip 13pt plus3pt minus3pt%
\belowdisplayskip \abovedisplayskip
\abovedisplayshortskip \z@ plus3pt%
\belowdisplayshortskip 7pt plus3.5pt minus0pt
\def\@listi{\parsep 4.5pt plus 2pt minus 1pt
      \itemsep \parsep
      \topsep 9pt plus 3pt minus 3pt}}

\relax

\catcode`@=12

\catcode`\@=11

\def\section{\@startsection{section}{1}{\z@}{3.5ex plus 1ex minus
    .2ex}{2.3ex plus .2ex}{\large\bf}}

\def\thesection{\arabic{section}}
\def\thesubsection{\arabic{section}.\arabic{subsection}}

\def\appendix{\setcounter{section}{0}
  \def\thesection{Appendix \Alph{section}}
  \def\thesubsection{\Alph{section}.\arabic{subsection}}
  \def\theequation{\Alph{section}.\arabic{equation}}}
\def\SymBoxes#1#2#3#4{\newdimen\un@t \un@t#3%
\raisebox{#1}{\rule{#2\un@t}{#4}\hskip-#2\un@t% lower horizontal
\@tempdimb\un@t \advance\@tempdimb by-#4\@tempcntb#2\relax%
\@whilenum{\@tempcntb>0}\do{%                         % #2 vertical lines
\rule{#4}{\un@t}\hskip\@tempdimb \advance\@tempcntb by\m@ne}%
\hskip-#2\un@t \rule[\un@t]{#2\un@t}{#4}%
\rule[\un@t]{#4}{#4}\hskip-#4%             % upper horizontal line
\rule{#4}{\un@t}}\hskip-#4}                % rightest vertical line
\begin{document}
%\begin{letter}{~}

%%%%%%Define some new commands and  macros

\newcommand{\dd}{\textrm{d}}

\newcommand{\beq}{\begin{equation}}
\newcommand{\eeq}{\end{equation}}
\newcommand{\bea}{\begin{eqnarray}}
\newcommand{\eea}{\end{eqnarray}}
\newcommand{\beas}{\begin{eqnarray*}}
\newcommand{\eeas}{\end{eqnarray*}}
\newcommand{\defi}{\stackrel{\rm def}{=}}
\newcommand{\non}{\nonumber}
\newcommand{\bquo}{\begin{quote}}
\newcommand{\enqu}{\end{quote}}
\newcommand{\tc}[1]{\textcolor{blue}{#1}}
%%%%%%%%%%%%%%%%
\renewcommand{\(}{\begin{equation}}
\renewcommand{\)}{\end{equation}}
%%%%%%%%%%%%%%%%%%%%%%%%%%%%%%%%%% definitions
\def\de{\partial}
\def\Om{\ensuremath{\Omega}}
\def\Tr{ \hbox{\rm Tr}}
\def\rc{ \hbox{$r_{\rm c}$}}
\def\H{ \hbox{\rm H}}
\def\HE{ \hbox{$\rm H^{even}$}}
\def\HO{ \hbox{$\rm H^{odd}$}}
\def\HEO{ \hbox{$\rm H^{even/odd}$}}
\def\HOE{ \hbox{$\rm H^{odd/even}$}}
\def\HHO{ \hbox{$\rm H_H^{odd}$}}
\def\HHEO{ \hbox{$\rm H_H^{even/odd}$}}
\def\HHOE{ \hbox{$\rm H_H^{odd/even}$}}
\def\K{ \hbox{\rm K}}
\def\Im{ \hbox{\rm Im}}
\def\Ker{ \hbox{\rm Ker}}
\def\const{\hbox {\rm const.}}
\def\o{\over}
\def\im{\hbox{\rm Im}}
\def\re{\hbox{\rm Re}}
\def\bra{\langle}\def\ket{\rangle}
\def\Arg{\hbox {\rm Arg}}
\def\exo{\hbox {\rm exp}}
\def\diag{\hbox{\rm diag}}
\def\longvert{{\rule[-2mm]{0.1mm}{7mm}}\,}
\def\a{\alpha}
\def\b{\beta}
\def\e{\epsilon}
\def\l{\lambda}
\def\ol{{\overline{\lambda}}}
\def\ochi{{\overline{\chi}}}
\def\th{\theta}
\def\s{\sigma}
\def\oth{\overline{\theta}}
\def\ad{{\dot{\alpha}}}
\def\bd{{\dot{\beta}}}
\def\oD{\overline{D}}
\def\opsi{\overline{\psi}}
\def\dag{{}^{\dagger}}
\def\tq{{\widetilde q}}
\def\L{{\mathcal{L}}}
\def\p{{}^{\prime}}
\def\W{W}
\def\N{{\cal N}}
\def\hsp{,\hspace{.7cm}}
\def\hspp{,\hspace{.5cm}}
\def\bo{\ensuremath{\hat{b}_1}}
\def\bfo{\ensuremath{\hat{b}_4}}
\def\co{\ensuremath{\hat{c}_1}}
\def\cfo{\ensuremath{\hat{c}_4}}
\def\th#1#2{\ensuremath{\theta_{#1#2}}}
\def\c#1#2{\hbox{\rm cos}(\th#1#2)}
\def\s#1#2{\hbox{\rm sin}(\th#1#2)}
\def\cp#1#2#3{\hbox{\rm cos}^#1(\th#2#3)}
\def\sp#1#2#3{\hbox{\rm sin}^#1(\th#2#3)}
\def\ctp#1#2#3{\hbox{\rm cot}^#1(\th#2#3)}
\def\cpp#1#2#3#4{\hbox{\rm cos}^#1(#2\th#3#4)}
\def\spp#1#2#3#4{\hbox{\rm sin}^#1(#2\th#3#4)}
\def\t#1#2{\hbox{\rm tan}(\th#1#2)}
\def\tp#1#2#3{\hbox{\rm tan}^#1(\th#2#3)}
\def\m#1#2{\ensuremath{\Delta M_{#1#2}^2}}
\def\mn#1#2{\ensuremath{|\Delta M_{#1#2}^2}|}
\def\u#1#2{\ensuremath{{}^{2#1#2}\mathrm{U}}}
\def\pu#1#2{\ensuremath{{}^{2#1#2}\mathrm{Pu}}}
\def\meff{\ensuremath{\Delta M^2_{\rm{eff}}}}
\def\an{\ensuremath{\alpha_n}}
\newcommand{\Z}{\ensuremath{\mathbb Z}}
\newcommand{\R}{\ensuremath{\mathbb R}}
\newcommand{\rp}{\ensuremath{\mathbb {RP}}}
\newcommand{\vac}{\ensuremath{|0\rangle}}
\newcommand{\vact}{\ensuremath{|00\rangle}                    }
\newcommand{\oc}{\ensuremath{\overline{c}}}
\renewcommand{\cos}{\textrm{cos}}
\renewcommand{\sec}{\textrm{sec}}
\renewcommand{\sin}{\textrm{sin}}
\renewcommand{\cot}{\textrm{cot}}
\renewcommand{\tan}{\textrm{tan}}
\renewcommand{\ln}{\textrm{ln}}

\newcommand{\Vol}{\textrm{Vol}}

\newcommand{\half}{\frac{1}{2}}

%%%%%%%%%%%%%%%%%%%%%%%Changed%%%%%%%%%%%%%%%%%%%%%%%%%%%%%
\def\changed#1{{\bf #1}}
%\def\changed#1{ #1}
%%%%%%%%%%%%%%%%%%%%%%%%%%%%%%%%%%%%%%%%%%%%%%%%%%%%%%%%%

\begin{titlepage}
\begin{flushright}
KEK-TH-1809\\
IFUP-TH/2015
\end{flushright}
\bigskip

\def\thefootnote{\fnsymbol{footnote}}

\begin{center}
{\large {\bf
The Leptonic CP Phase from T2(H)K and $\mu^+$ Decay at Rest
  } }
%\end{center}

\bigskip

\bigskip

{\large \noindent   Jarah Evslin$^{1,2,3}$\footnote{\texttt{jarah@impcas.ac.cn}},
Shao-Feng Ge${}^4$\footnote{\texttt{gesf02@gmail.com}} and 
Kaoru Hagiwara${}^{2,5,6}$\footnote{\texttt{kaoru.hagiwara@kek.jp}}}

%, Emilio Ciuffoli$^{3,5}$\footnote{ciuffoli@ihep.ac.cn} and Xinmin Zhang$^{3,5}$\footnote{\texttt{xmzhang@ihep.ac.cn}} }
\end{center}

\renewcommand{\thefootnote}{\arabic{footnote}}

\vskip.7cm

\begin{center}
\vspace{0em} {\em  
1)  Institute of Modern Physics, CAS, NanChangLu 509, Lanzhou 730000, China\\
2) KEK Theory Center, 1-1 Oho, Tsukuba, Ibaraki 305-0801, Japan\\
3) INFN Sezione di Pisa, Largo Pontecorvo 3, 56127, Pisa, Italy\\
4) Max-Planck-Institut f\"ur Kernphysik, Heidelberg 69117, Germany\\
5) Sokendai, 1-1 Oho, Tsukuba, Ibaraki 305-0801, Japan\\
6)  Department of Physics, University of Wisconsin-Madison,
  Madison, WI 53706, USA
\\}

%\vspace{0em} {\em  { 1) TPCSF, IHEP, Chinese Acad. of Sciences\\
%2) Theoretical physics division, IHEP, Chinese Acad. of Sciences\\
%YuQuan Lu 19(B), Beijing 100049, China}}

\vskip .4cm

\vskip .4cm

\end{center}

\vspace{1.3cm}

\noindent
\begin{center} {\bf Abstract} \end{center}

\noindent
Combining $\nu$ oscillations at T2K or T2HK with $\overline{\nu}$ oscillations from $\mu^+$ decay at rest (DAR) allows a determination of the leptonic CP-violating phase $\delta$.  The degeneracies of this phase with $\theta_{13}$ and $\theta_{23}$ are broken and $\delta$ can be reliably distinguished from $180^\circ-\delta$.  We present the sensitivity to $\delta$ of T2(H)K together with a $\mu^+$ DAR experiment using Super-K as a near detector and Hyper-K at the Tochibora site as a far detector.

%We present the sensitivity to $\delta$ of one $\mu^+$ DAR experiment with a single $\mu^+$ source using two JUNO or RENO-50 detectors and another using Super-K as a near detector and Hyper-K at the Tochibora site as a far detector.
\vfill

\begin{flushleft}
{\today}
%\vspace{1cm}
\end{flushleft}
\end{titlepage}
%\bigskip

\hfill{}
%\bigskip

%\tableofcontents

\setcounter{footnote}{0}

\section{Introduction} \label{intro}
\noindent
We propose the experiment Tokai 'N Toyama to Kamioka (TNT2K).   The 50 kton water Cherenkov detector Super-Kamiokande (SK)  will detect both $\nu_e$ appearance in a $\nu_\mu$ beam created at J-PARC in Tokai and also $\overline{\nu}_e$ appearance in isotropic $\overline{\nu}_\mu$ created by $\mu^+$ decay at rest ($\mu$DAR) at a high intensity accelerator just south of Toyama.  We will show that this yields a precise determination of the leptonic CP-violating phase $\delta$.  With the addition of just one fifth of Hyper-Kamiokande (HK) at the preferred Tochibora site, we find that $\delta$ can be determined more precisely than with T2HK using the full megaton HK but no $\mu$DAR and also $\delta$ can be reliably distinguished from $180^\circ-\delta$.

The J-PARC beam creates $\nu_\mu$ by colliding a  750 kW, 30 GeV proton beam into a target, ejecting $\pi^+$ which are filtered into a tunnel where they decay into a $\mu^+$ and $\nu_\mu$.   As $\overline{\nu}$ oscillations are provided by $\mu$DAR from the Toyama accelerator, an optimal determination of $\delta$ arises when the J-PARC beam runs exclusively in $\nu$ mode.  The $\nu_\mu$ travel along the beam, oscillating as they go, and SK detects, largely via charged current quasielastic (CCQE) interactions, both $\nu_e$ appearance in the beam and also the $\nu_\mu$ disappearance.   SK is located off of the beam axis, where the beam is relatively monochromatic, centered on the first oscillation maximum which at 295 km is at about 600 MeV.  The Tochibora HK site is at the same baseline and off-axis angle, and so if built will also observe $\nu$ at the first oscillation maximum.

\begin{figure} %[!tph]
\begin{center}
\includegraphics[angle=-0,width=2.2in]{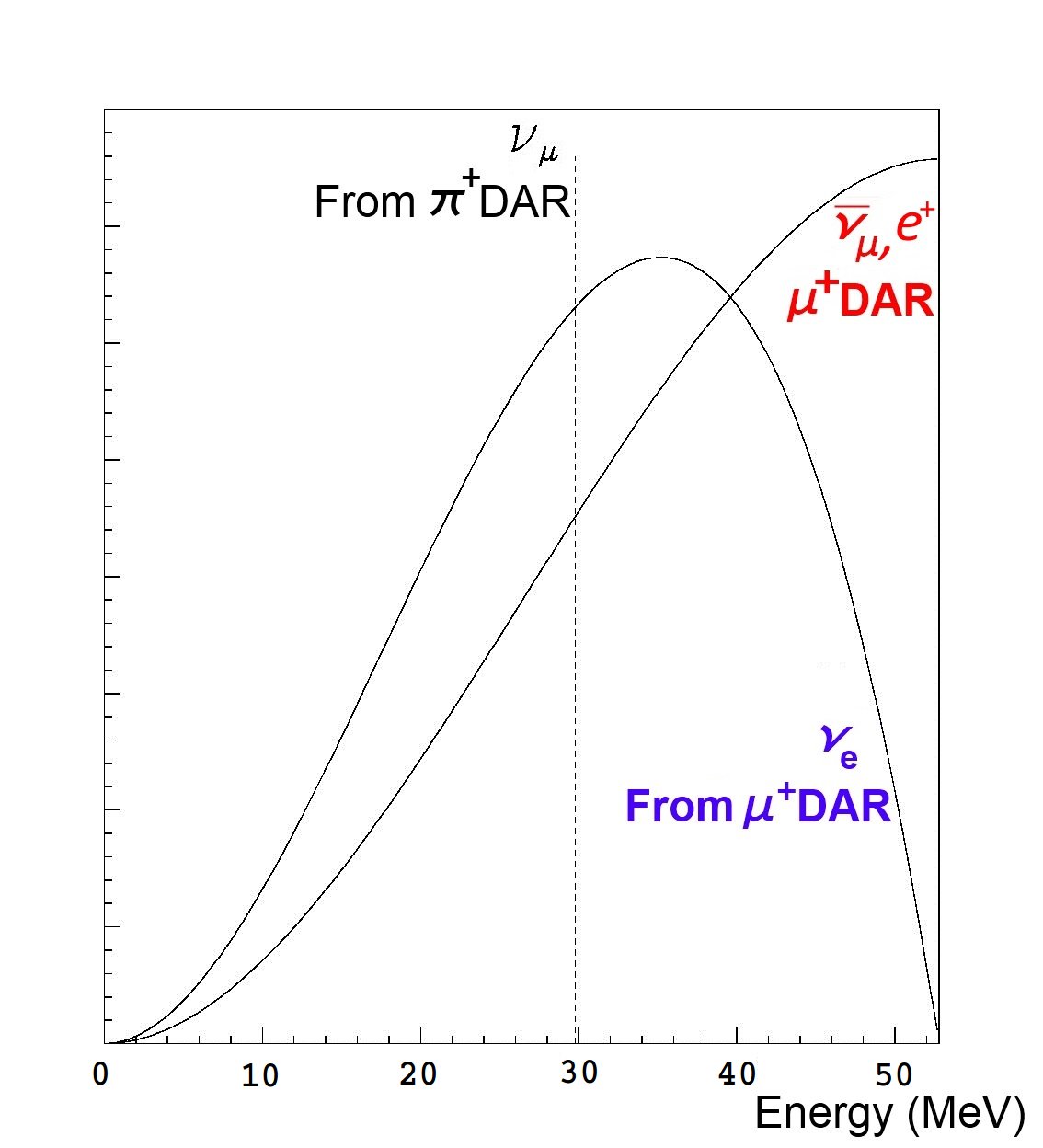}
\caption{The neutrino spectrum produced by $\pi^+$ and $\mu^+$ decay at rest (DAR).  The charge conjugates are produced by $\pi^-$ and $\mu^-$ decay at rest, which we assume to be suppressed by a factor of $5\times 10^{-4}$.}   
\label{michelafig}
\end{center}
\end{figure}

The $\mu$DAR source collides protons into a target, creating $\pi^+$ and $\pi^-$.  The proton beam must be of low enough energy, and the target large enough, so that the resulting $\pi$ stop in the target.  The $\pi^-$ are absorbed or decay into $\overline{\nu}_\mu$ and $\mu^-$ which are absorbed in a sufficiently high $Z$ target.  On the other hand the $\pi^+$ decay at rest, yielding $\mu^+$ and $\nu_\mu$.  The $\mu^+$ in turn also stop and decay at rest, producing $e^+$, $\overline{\nu}_\mu$ and $\nu_e$.  The spectra of all of these neutrinos are illustrated in Fig.~\ref{michelafig}.  The experiment searches for the conversion of $\overline{\nu}_\mu$ to $\overline{\nu}_e$ between the source and the detector(s).  The $\mu^+$ decay at rest spectrum is known quite precisely.  Most of the $\overline{\nu}_e$ will have energies of between 30 MeV and 50 MeV and so will interact with SK and HK via inverse $\beta$ decay (IBD), whose cross section is also known quite precisely.   IBD creates an additional neutron whose capture SK-IV is sometimes able to detect \cite{skwendell} and use to reduce backgrounds.  It is expected that, despite its lower PMT coverage, HK will have the same ability.  However, to be conservative, we do not use this in our analysis.

SK and HK can easily distinguish the low energy $\mu$DAR $\overline{\nu}_e$ from pulsed, higher energy J-PARC $\nu$ and so the $\mu$DAR and beam $\nu$ experiments can run simultaneously.  We will show that the optimal location for the $\mu$DAR source is 15 km north of SK, which is 23 km north of HK, in the southern hills of Toyama.  Fortunately the optimal distance to SK is roughly the same while HK is not in operation.  Even if HK is not constructed, with a combination of $\mu$DAR and the T2K $\nu$ beam one can determine $\delta$ with moderate accuracy.       The favored location for HK is the Tochibora mine, 8 km south of SK \cite{HK,HK2}.  However, a site in the Mozumi mine has also been considered.  As this location is very close to SK, if it is chosen then the TNT2K experiment will be less precise, suffering from the $\delta\leftrightarrow 180^\circ-\delta$ degeneracy.

% independent of whether or not HK is constructed.  

The Toyama accelerator needs to accelerate protons to between 400 MeV and 3 GeV, with an optimal performance per MW between 600 MeV and 1.5 GeV.  So far the most powerful such accelerator is the 2.2 mA, 600 MeV proton cyclotron at the Paul Scherrer Institute.   As explained in Ref. \cite{atomic}, an optimization of that design would allow for a 5 mA, 800 MeV proton beam, which would already be sufficient for our purposes.   The target station in, for example, Ref.~\cite{graphite} would be suitable.  On the other hand, there are currently efforts around the world to run accelerator driven subcritical reactors.  For example, the project \cite{ads} envisages  a 10 mA, 1.5 GW beam with an intermediate stage which is already 10 mA at 800 MeV.    For concreteness we will choose intermediate parameters, a 9 mA, 800 MeV beam which could be achieved for example with two of the accelerators of Ref.~\cite{atomic}.   

In order to determine the $\overline{\nu}_\mu$ flux normalization we also require a small, near detector.  For example a 20 ton liquid scintillator detector, such as one of the eight which Daya Bay will no longer need or perhaps one from RENO, would be quite sufficient.  Using elastic scattering, such a detector should be able to determine the flux normalization to within about 5\% \cite{lsnd97} and also to provide a very powerful check of the LSND anomaly \cite{lsnd01} with a reach to lower mixings and mass splittings than LSND itself.

At the first oscillation maximum,  the uncertainty in $\spp2213\sp223$ is about half as large as the maximal contribution of $\sin(\delta)$ to the  $\nu_\mu\rightarrow\nu_e$ oscillation probability $P_{\mu e}$ and so $\nu_\mu\rightarrow\nu_e$ oscillations alone cannot demonstrate leptonic CP-violating beyond the 2$\sigma$ level.  This problem can be resolved by combining $\nu_\mu\rightarrow\nu_e$ and $\overline{\nu}_\mu\rightarrow\overline{\nu}_e$ oscillations because, at the first oscillation maximum, the sum of the oscillation probabilities $P_{\mu e}+P_{\overline{\mu}\overline{e}}$ depends upon $\spp2213\sp223$ and to a lesser extent on $\cos(\delta)$ while the difference $P_{\mu e}-P_{\overline{\mu}\overline{e}}$ depends upon $\sin(\delta)$.  Thus by comparing the $\nu$ beam and $\mu$DAR $\overline{\nu}$ experiments one can accurately extract $\sin(\delta)$, which measures leptonic CP-violation. 

 On the other hand, the existence of multiple baselines is useful not only to control systematic errors \cite{daedwhite} but also to extract $\cos(\delta)$ \cite{noidarts}, thus breaking the $\delta\rightarrow180^\circ-\delta$ degeneracy present in MINOS, T2K, NO$\nu$A, MOMENT and other beam experiments \cite{noiinterf,parkedegen}.    Motivation for measuring $\cos(\delta)$ is given in Refs.~\cite{shaofeng1,shaofeng2,peter1,shaofeng3,petcos1,peter2,petcos2,titov}.

While our proposal for the measurement of $\delta$ is in spirit similar to that of the DAE$\delta$ALUS project \cite{daed,daedwhite}, it differs in one key respect.  DAE$\delta$ALUS uses 3 cyclotron complexes as $\mu$DAR sources.  However, as the direction of the IBD positron is only very weakly correlated with that of the incoming $\overline{\nu}_e$, the spectra from the different sources can only be separated by running just one accelerator at a time.  Thus each runs with a duty factor of only 20\% and so requires an extremely high instantaneous intensity.  To achieve this high intensity it will accelerate $H_2^+$ molecules but this involves technological progress, for example, the excited molecules must be removed.  Our proposal is designed to be cheaper because only a single cyclotron complex is necessary and, as it may in principle run continuously, the instantaneous intensity may be reduced by up to a factor of five.    

The 20\% duty factor at DAE$\delta$ALUS serves not only so that only one cyclotron runs at a time, but also the 40\% dead time allows one to measure backgrounds.  However, as will be explained below, by far the dominant background at our $\mu$DAR experiment arises from invisible muons created by atmospheric neutrinos.  These lead to a background with a known shape, so only the normalization must be determined.  However SK has been measuring this background, as part of its diffuse supernova neutrino search, for nearly 20 years \cite{sksnold,sksn,sksn3}.  Thus, while some accelerator downtime is inevitable and this will be used to measure the atmospheric backgrounds, which are after all dependent on the season and solar activity, we do not require a structured beam.

In this regard the locations of our experiment in western Japan provide yet another advantage.  The main backgrounds arise from low energy atmospheric neutrinos.    According to the model of Ref. \cite{campimag} at the Kamioka mines the horizontal component of the geomagnetic field is 0.31 Gauss, appreciably higher than the 0.17 Gauss that may be expected at DUNE or the 0.13 Gauss at LENA in the Pyh\"asalmi mine.  As this strong horizontal field deflects low energy cosmic rays, the atmospheric neutrino backgrounds at the sites suggested in our proposal will be reduced by nearly one half \cite{atm1989,honda} with respect to the other sites at which $\mu^+$ DAR measurements of $\delta$ have been proposed.

\section{Parameters}

\subsection{The Neutrino Mass Matrix} \label{angoli}

We fix the solar neutrino mass splitting to be
\beq
%\m31=2.4\times 10^{-3}{\rm{eV}}^2\hsp
\m21=7.5\pm 0.2\times 10^{-5}{\rm{eV}}^2
\eeq
and, for ease of comparison with previous studies, we choose the neutrino mass matrix mixing angles to be
\beq
\spp2213=0.089,\ 
\spp2212=0.857,\ 
\sp223=\frac{1}{2}.
\eeq
The current uncertainties \cite{pdg,dayaboston,t2kdisp2014}
\bea
&&\delta\spp2212=0.024\hsp \delta\spp2213=0.005\nonumber\\
&& \delta\sin(\theta_{23})=0.055 
\eea
are used.
%For simplicity we have assumed a simple Gaussian error for each mixing angle, even $\theta_{23}$, despite the fact that nonmaximal mixing can imply that measurements of $\sp223$ yield two disconnected confidence intervals for this angle, one in each octant.  We feel that this is reasonable because T2K and T2HK are instead sensitive to $\sin(\theta_{23})$ and so are likely to break the octant degeneracy when the mixing is sufficiently nonmaximal for it to be relevant.  More to the point, their appearance channel results will provide strong bounds directly upon virtually the same combination of $\theta_{23}$ and $\theta_{13}$ which is degenerate with $\delta$ in DAR experiments.

We assume that the neutrino mass hierarchy has already been determined when our experiment has collected its data.  While the choice of the hierarchy has little effect on the sensitivity to $\delta$, the assumption that it is known does break a degeneracy in experiments, such as T2K, NO$\nu$A and T2HK, in which the matter effect is appreciable.  As the experiments proposed here are unlikely to be performed in the next 10 years, it is reasonable to assume that the hierarchy will be known with some cautious certainty.

Currently, the atmospheric mass splitting $\Delta M^2_{\mu\mu}$ \cite{parke2005} has only been measured at the 4\% level \cite{minosmassa,t2kdisp2014}.  Daya Bay has matched this precision \cite{dayam} for the corresponding effective mass splitting $\Delta M^2_{ee}$ \cite{parke2005}.  However, the disappearance channels at T2K and NO$\nu$A will each achieve a better than 2\% precision, with a 1\% precision possible when they are combined \cite{weit2k}, while JUNO and RENO-50 are each expected to achieve a subpercent precision \cite{shaofengprec}.  We will consider the uncertainty of the MINOS measurement \cite{minosmassa} together with the hierarchy-averaged central value%which in the case of the normal hierarchy is
\beq
\m31=(2.4\pm 0.1)\times 10^{-3}{\rm{eV}}^2.
%\m31=(2.35\pm 0.11)\times 10^{-3}{\rm{eV}}^2.
\eeq

%In Fig. \ref{pfig} the mass matrix parameters described here are used to determine the probability of the oscillation $\overline{\nu}_\mu\rightarrow\overline{\nu}_e$ at the baselines of interest for these proposals.  The $\nu_\mu\rightarrow\nu_e$ and $\nu_\mu$ survival probabilities relevant to T2K and T2HK are summarized in Fig \ref{t2kpfig}.  Only the values $\delta=0^\circ,\ 90^\circ,\ 180^\circ$ and $270^\circ$ are displayed.

Recently SK has reported an excess of $\nu_\mu\rightarrow\nu_e$ in the J-PARC beam \cite{t2kexcess} and also a low energy atmospheric $\nu_e$ excess corresponding to a deficit in $\nu_e\rightarrow\nu_\mu$ \cite{skwendell}, which together give roughly a 2$\sigma$ preference to $\delta=240^\circ$ over $\delta=60^\circ$, although null CP violation is allowed within 1$\sigma$.    On the other hand, a small deficit in accelerator neutrino $\nu_\mu\rightarrow\nu_e$ at MINOS has led to a statistically insignificant preference for $\sin(\delta)>0$ \cite{minoscp}.  In this note we will not consider any of these hints in our analysis.

\subsection{Experimental Setup}

We normalize the detector efficiency and the $\mu^+$ DAR rate such that, at 10 km if $\delta=0$, 350 inverse $\beta$ decay events, corresponding to $\overline{\nu}_e$ capture on free protons in SK would be observed in a 6 year run.  By scaling results from LSND \cite{lsnd01} this roughly corresponds to 6 years of collisions of an $800$ MeV proton beam on a stationary target if the beam current is 9 mA.  The beam is not pulsed:  although clearly a real beam will have dead time which can serve to measure the background, we approximate our duty factor to be 100\%.  Thus the maximum instantaneous current is also 9 mA, a factor of 4 less than that which will be required at DAE$\delta$ALUS \cite{daedseminario}. The integrated current corresponds to a total of $1.1\times 10^{25}$ protons on target (POT), 60 times more than LSND.  %The uncertainty on the total number of events will be taken to be 5\%. 

\begin{figure} %[!tph]
\begin{center}
\includegraphics[angle=-0,width=3.2in]{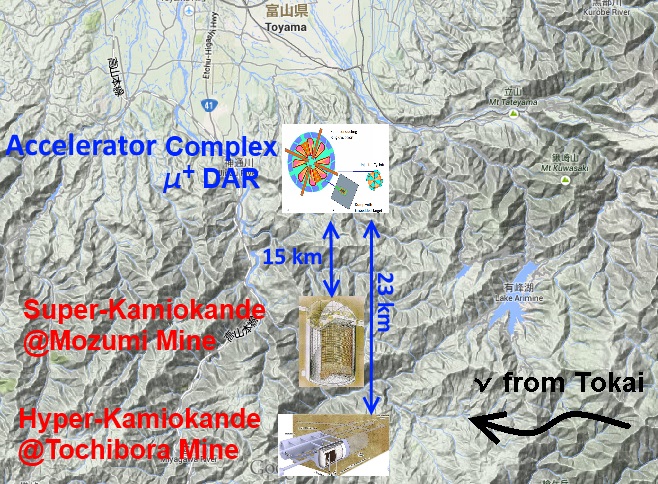}
\caption{Hyper-K and Super-K will be the near and far detectors for $\mu^+$DAR $\overline{\nu}$ from an accelerator complex just south of Toyama, while simultaneously detecting higher energy $\nu$ from the J-PARC beam.}   
\label{mappafig}
\end{center}
\end{figure}

As SK and the Tochibora HK site are only separated by 8 km, the difference in the two baselines can at most be 8 km.  For the TNT2K experiment we will place the $\mu^+$ source 15~km north of SK, just south of Toyama city, as is illustrated in Fig.~\ref{mappafig}.  Thus the near and far baselines will be 15 and 23 km, respectively.  We will assume that J-PARC offers 750 kW of its beam to this effort, in line with the goal in the next 5 years stated in KEK's most recent road map \cite{5anni}.  Note that this is less than half of the beam power traditionally considered in simulations of T2HK \cite{HK}.

%Again the $\mu^+$ source will be a cyclotron complex with a 9 mA current of 800 MeV protons colliding with a fixed target.  %In keeping with Ref. \cite{HK} we assume that the uncertainty in the number of events from each experiment is 5\%.    

We consider SK and HK without gadolinium \cite{gadzooks}.  As a result of statistical fluctuations in the number of photoelectons, in the energy range relevant to the DAR experiments we will consider fractional energy resolutions of 
\beq
\frac{\delta E}{E}=\frac{40\%}{\sqrt{E/{\rm{MeV}}}}\rm{\ and\ }\frac{60\%}{\sqrt{E/{\rm{MeV}}}} \label{enres}
\eeq
respectively. 

In the case of higher energy accelerator neutrinos from J-PARC, the energy resolutions are no longer limited entirely by photoelectron statistics, and so are somewhat worse than one would extrapolate from Eq. (\ref{enres}) as can be seen in Refs. \cite{SK2005} and \cite{HK} for SK and HK respectively.  In addition, $\Delta$ resonance charged current interactions transfer some of the neutrino energy into additional pions and so yield an average energy which is reduced by about 360 MeV \cite{kaorusk}.  We incorporate the reduction in energy of some events and the energy resolution by folding the true spectrum with the sum of three Gaussians whose forms are given in Appendix A of Ref. \cite{kaorusk}.

% we use instead momentum resolutions of \cite{SK2005}
%\beq
%\frac{\delta p}{p}=6\times 10^{-3}+\frac{0.026}{\sqrt{p/{\rm{GeV}}}}\rm{\ and\ }\frac{???}{\sqrt{p/{\rm{MeV}}}}
%\eeq
%at T2K and T2HK respectively for electrons and
%\beq
%\frac{\delta p}{p}=0.017+\frac{7\times 10^{-3}}{\sqrt{p/{\rm{GeV}}}}\rm{\ and\ }\frac{???}{\sqrt{p/{\rm{GeV}}}}
%\eeq
%for muons. 

\begin{figure} %[!tph]
\begin{center}
\includegraphics[angle=-0,width=3.2in]{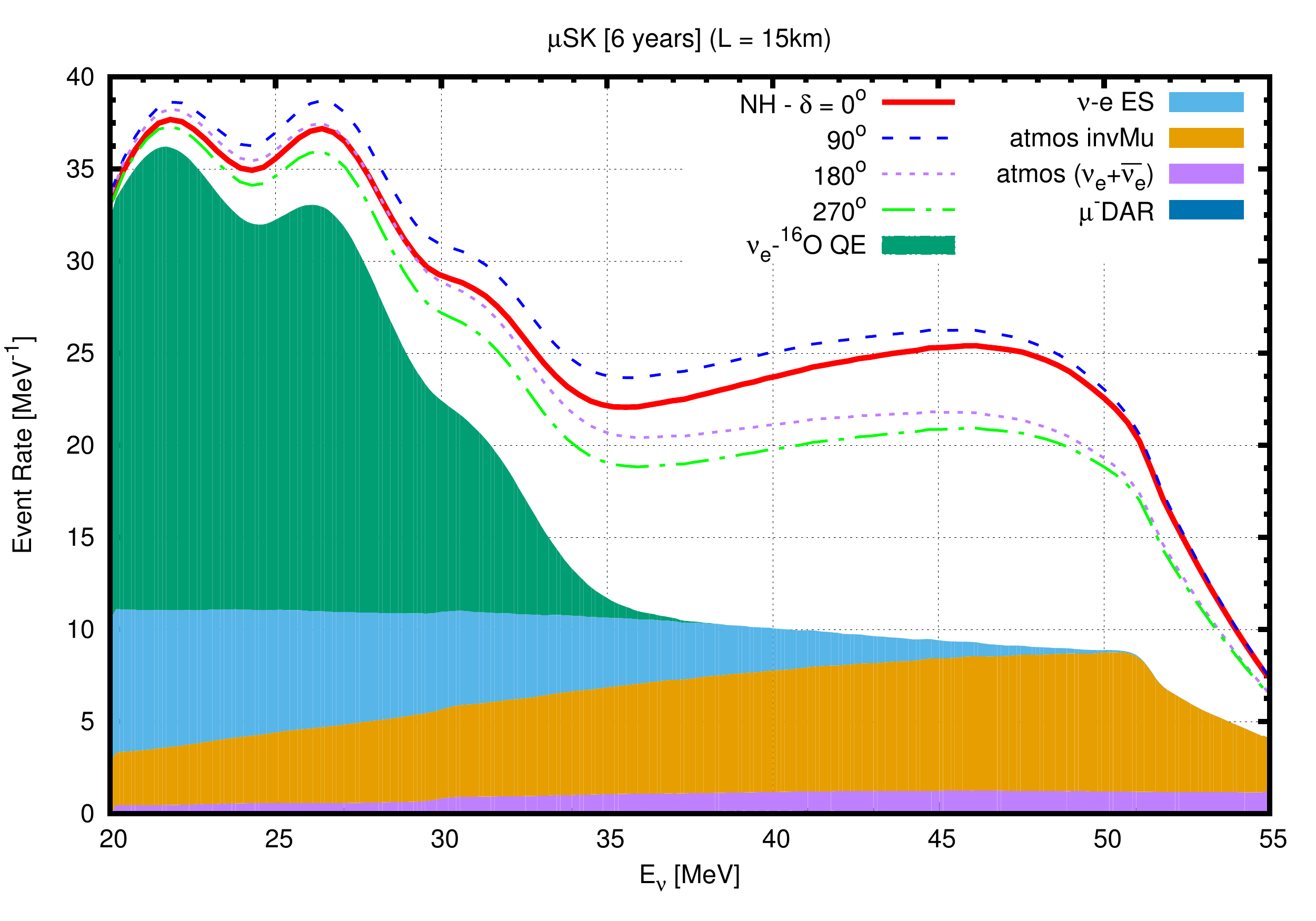}
\includegraphics[angle=-0,width=3.2in]{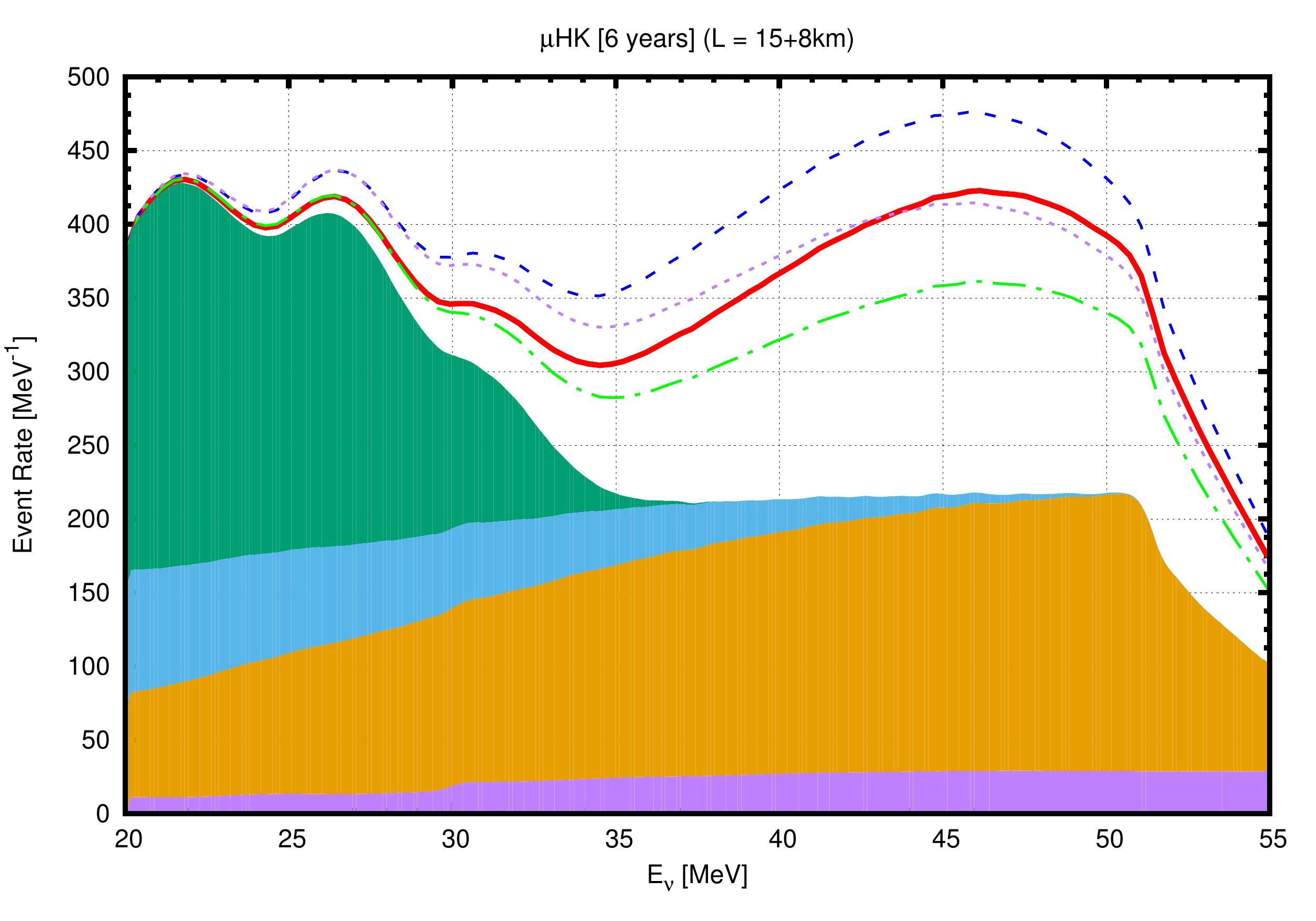}
\caption{The expected 12 and 6 year $\mu$DAR signals and backgrounds at the 15 km and 23 km of the TNT2K detectors SK and HK in the case of the normal hierarchy.}%  The total background and the electron (anti)neutrino background are included.}
\label{eventifig}
\end{center}
\end{figure}

\begin{figure} %[!tph]
\begin{center}
\includegraphics[angle=-0,width=3.2in]{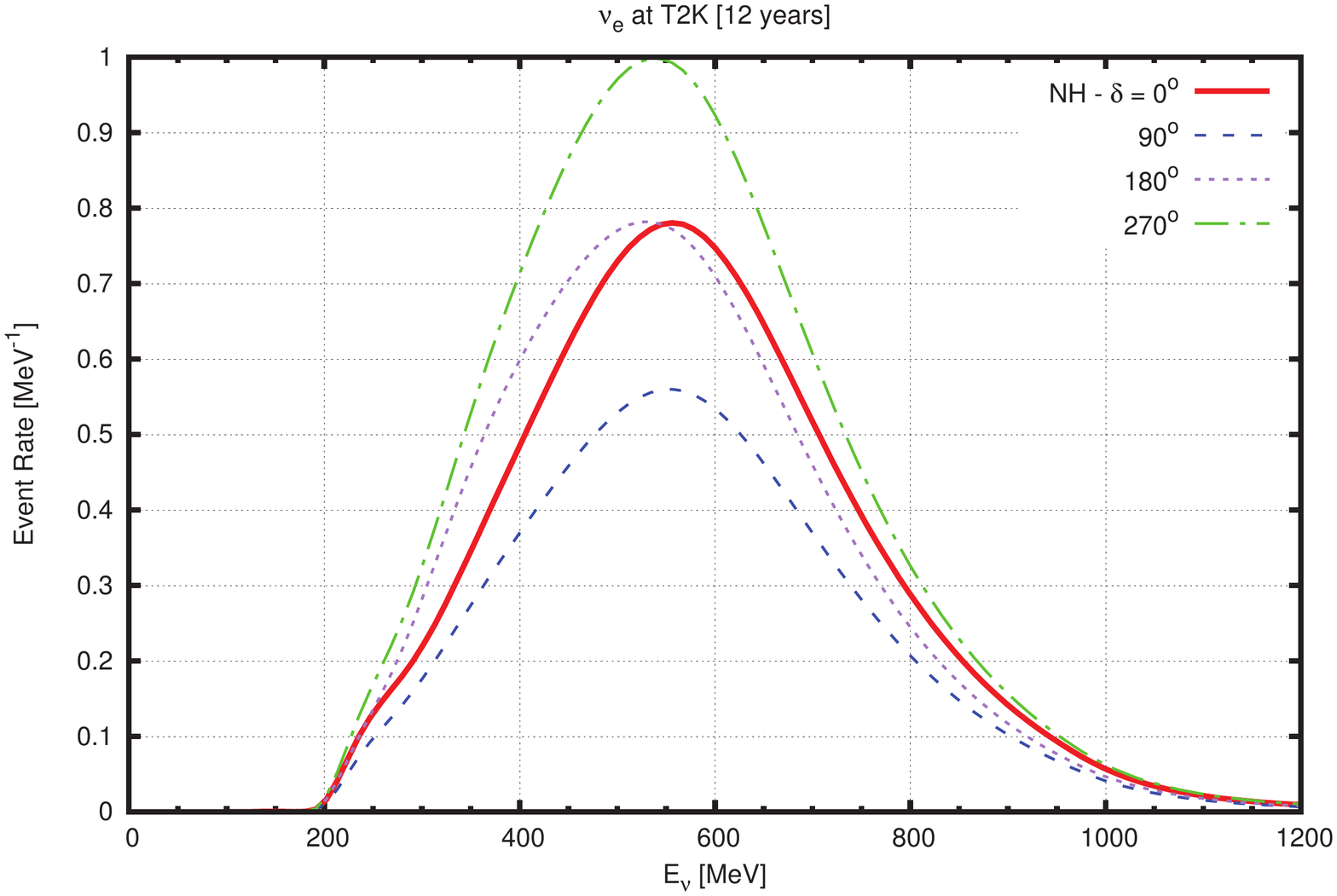}
\includegraphics[angle=-0,width=3.2in]{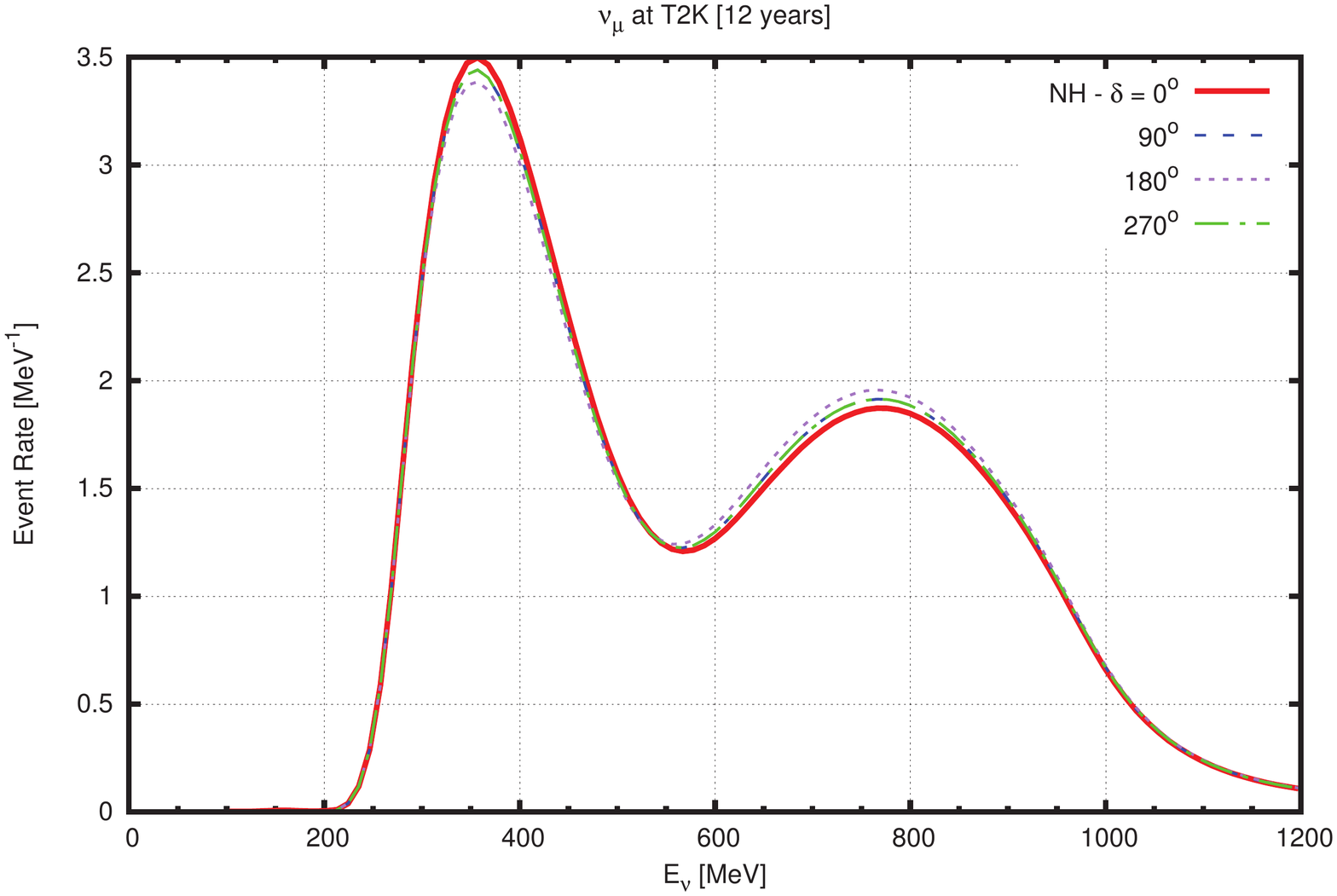}
\includegraphics[angle=-0,width=3.2in]{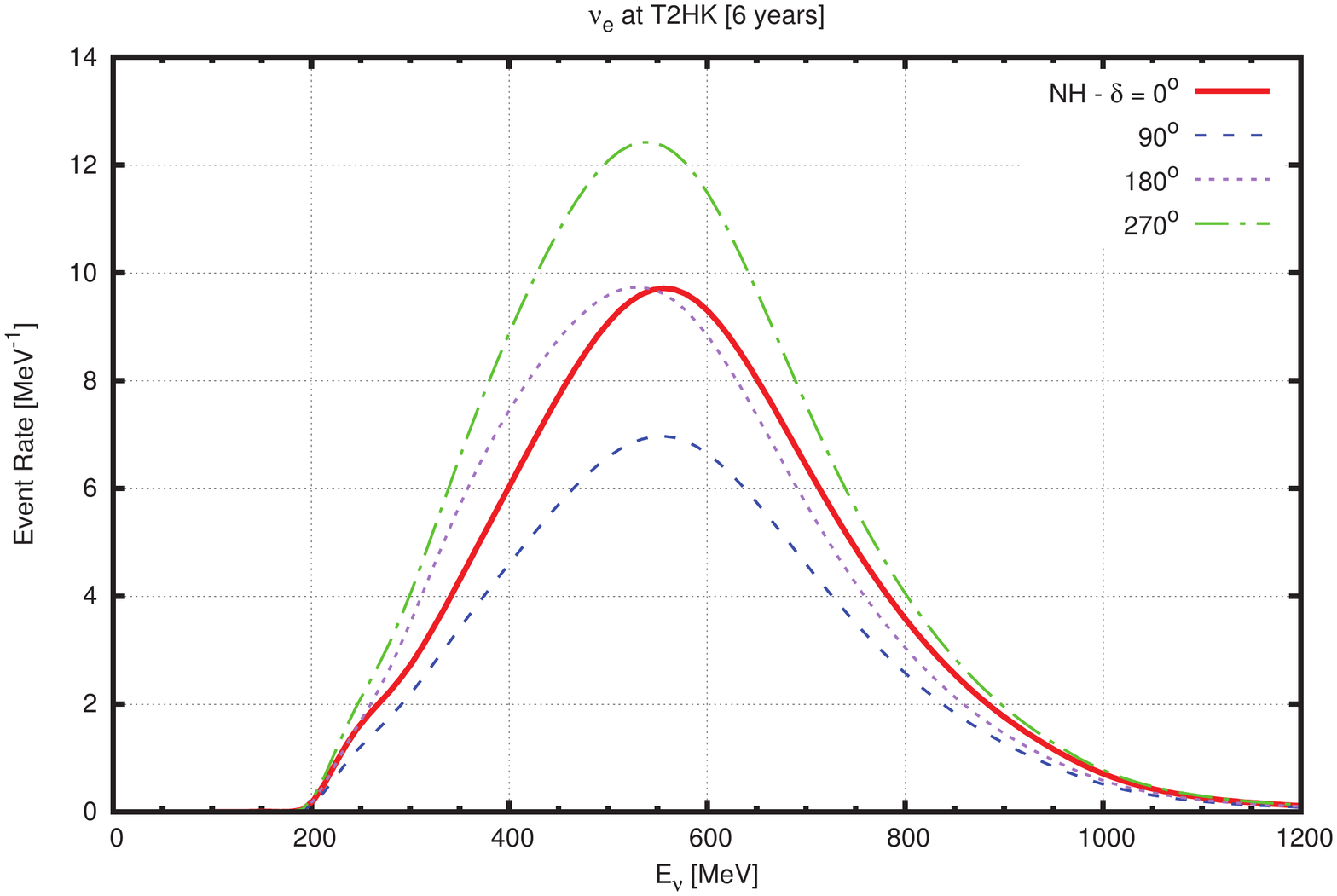}
\includegraphics[angle=-0,width=3.2in]{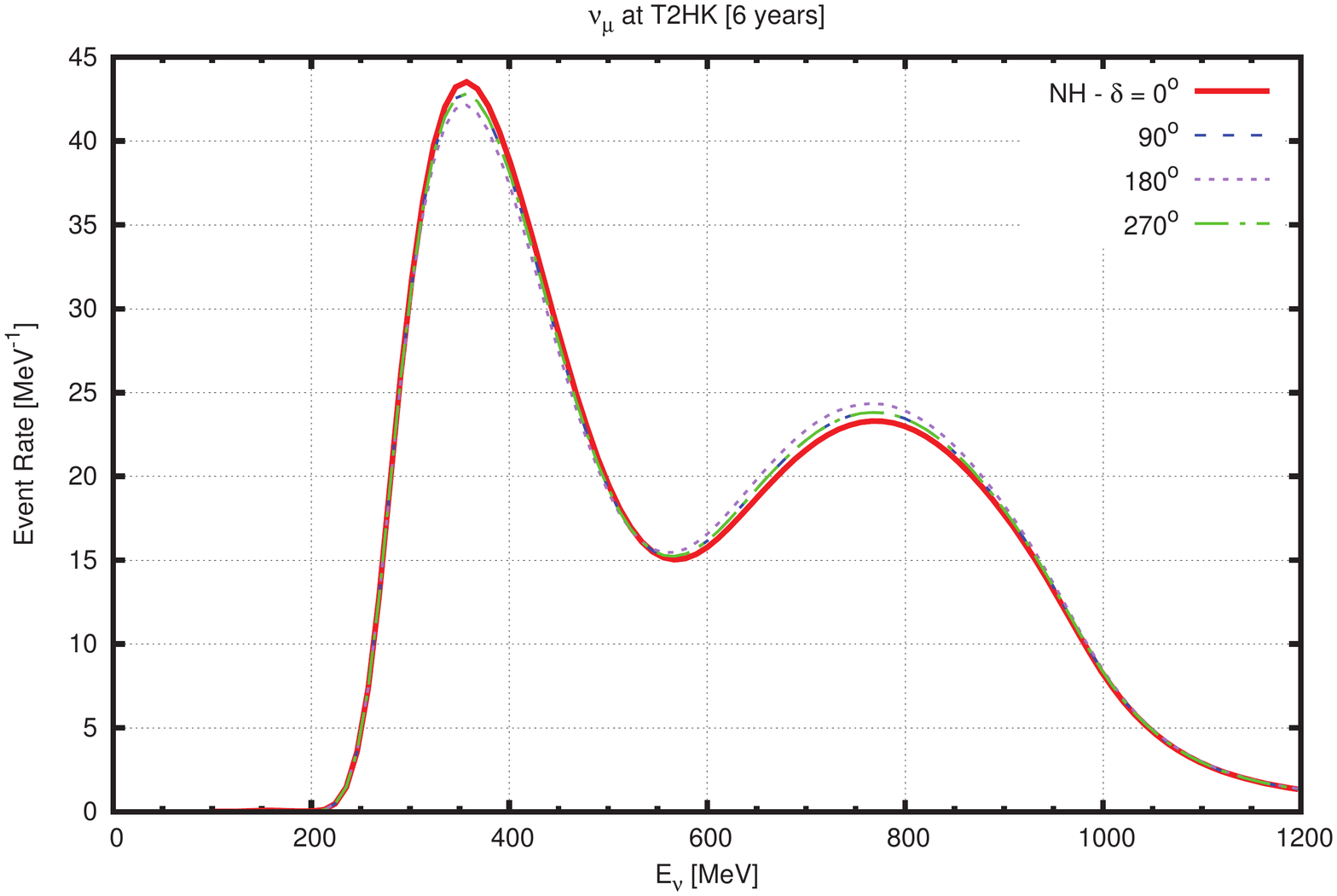}
\caption{The expected appearance and disappearance channel spectra at T2K (T2HK) after running for 6 (12) years in the $\nu$ mode.}   
\label{t2keventifig}
\end{center}
\end{figure}

The expected $\mu^+$ DAR signals are reported in Fig \ref{eventifig} while the expected appearance and disappearance spectra at T2K and T2HK are reported in Fig \ref{t2keventifig}.   These were obtained using the NuPro package \cite{nupro} and, in many cases, were confronted with the results of an independent C++ code \cite{emilio}.    We used definition of T2(H)K in Ref.~\cite{globest2k} with a target fiducial mass of 22.5 kton (560 kton) and a 750 kW beams operating $10^7$ seconds/year for 12 years (6 years) in $\nu$ mode.

\subsection{Flux uncertainty}

The uncertainty in the J-PARC $\nu$ and $\overline{\nu}$ rates (flux times cross-section) is taken to be 5\%, with the uncertainties uncorrelated.  On the other hand, we make a crude approximation that they are 100\% correlated between T2K and T2HK.

The uncertainty in the total DAR event rate reflects the uncertainty in the $\mu^+$ DAR rate itself at the target and also the efficiencies of the detector.   The former can be determined using dead reckoning with an accuracy of about 20\%.  As has been demonstrated by LSND \cite{lsnd01}, a relatively small water or mineral oil based liquid scintillator near detector has a good energy resolution in the relevant energy range and can determine the flux much more precisely via various channels, such as neutrino electron elastic scattering, which can be separated from CCQE interactions using the fact that the angular distribution of elastically scattered electrons is strongly forward peaked.  We will assume that a near detector is built and so, together with a calibration of the DAR, the event rate normalization error can be reduced to only 5\%.

One choice for a near detector may be the 50 ton liquid scintillator detector proposed in Refs. \cite{sterile,sterile2}.  The authors proposed that this detector be built 17 meters from a 0.33 mA, 3 GeV accelerator and search for $\overline\nu_\mu\rightarrow\overline\nu_e$ oscillations which would indicate sterile neutrinos.  As this baseline coincides with that of KARMEN, which observed no such oscillations, the high $\Delta M^2$ regime probed by this experiment is largely excluded.  The authors mention that at a later time it may be desirable to build a farther detector to search the lower $\Delta M^2$ regime, but to obtain a sufficient number of events such a detector would need to be much larger.  In addition, the 3 GeV beam energy is not optimal for a DAR experiment.  It produces less DAR events per unit of beam power than a 800 MeV proton beam and has higher backgrounds, although the backgrounds can be reduced using the time structure of the 3 GeV beam.

The present proposal would address both of these issues.  The 800 MeV proton beam discussed here has 27 times the current of the J-PARC beam of Refs. \cite{sterile,sterile2}, even after an extensive upgrade.  Therefore the near detector may be placed 50 meters from the accelerator instead of 17 m.  This is even longer than the baseline at LSND, and so would extend the reach in $\Delta M^2$ for a sterile neutrino search throughout the entire region suggested by LSND.  Even at 50 m the detector would see three times as many events as the original proposal at 17 m.  The improved statistics mean that the shape of the observed spectrum could be used to differentiate sterile neutrino oscillations from other potential signals.  Furthermore, by tripling the distance from the accelerator a number of accelerator-related backgrounds are greatly reduced, as well as low energy cosmogenic muon backgrounds if the detector is placed 50 meters underground.

\section{Backgrounds}

Our signal and background rates are summarized in Fig.~\ref{eventifig} and Table~\ref{eventitottab}.  The shape of the signal spectrum is well known, it is the spectrum of $\overline{\nu}_\mu$ from $\mu^+$ decay at rest.  It vanishes above its maximum at 53 MeV and is appreciable above 30 MeV.  Below 10 MeV reactor neutrino backgrounds hopelessly dominate it.  Below 20 MeV it will be dominated by the decays of spallation products created by cosmogenic muons, although a double coincidence with neutron capture could reduce this background considerably.  Near 20 MeV the signal may also be subdominant to the diffuse supernova background.   The regime from 30 MeV to 53 MeV will be the main focus of these experiments, containing the first oscillation peak of $\overline{\nu}_e$ appearance.  While the second oscillation peak is more sensitive to $\delta$, it would be extremely challenging to resolve above the backgrounds.

Our signal arises from inverse beta capture of oscillated $\overline{\nu}_e$, from $\mu^+$DAR, on free protons.  However the target volume also contains oxygen nuclei. These can interact via a quasielastic interaction with both the oscillated $\overline{\nu}_e$ and also the much more numerous unoscillated $\nu_e$ from $\mu^+$DAR.  The rate of the CCQE interaction
\beq
\nu_e+{}^{16}O\longrightarrow e^-+{}^{16}F
\eeq
is actually higher than our signal rate.  However the $Q$ value of 15.9 MeV, combined with the fairly low $\nu_e$ flux above 45 MeV, implies that only a small fraction of the electrons have energies above 29 MeV and so will be removed by our low energy veto.   This leaves the background shown in Fig.~\ref{eventifig}, which is only appreciable below 35 MeV.  It was calculated by folding the $\mu$DAR $\nu_e$ spectrum into GENIE and shifting the energy by hand to reproduce the correct $Q$ value, and renormalizing the event rate to agree with the calculated values in Ref.~\cite{volpe}.  Note that this background is much larger at DAE$\delta$ALUS \cite{daed}, where the low energy veto is 20 MeV.  For completeness, in Fig.~\ref{eventifig} we have also included the $\nu$-$e^-$ elastic scattering background, although the vast majority of these events can be removed using an angular veto \cite{lsnd97}.  

There are also several beam on backgrounds.  By far the largest of these arrives as follows.  When the beam hits the target, not only $\pi^+$ are made, but also $\pi^-$.  For example, at LSND the ratio of $\pi^-$ to $\pi^+$ is about 1 to 8 \cite{lsnd97}.  Most $\pi^-$ stop in the target and are immediately absorbed.  However some, about 5\% at LSND, decay in flight yielding $\mu^-$.  At least 90\% of $\mu^-$ are absorbed in a high $Z$ ($Z\geq 20$) target \cite{muonirev}.  The rest stop or are captured into orbit about oxygen nucleii.  Their decay, which we loosely refer to as $\mu^-$DAR, yields an irreducible background of $\overline{\nu}_e$ with a spectrum which is similar to our signal.  This background is small but difficult to quantify.  Roughly following the estimates above, we have fixed the ratio of $\mu^-$DAR to $\mu^+$DAR to be $5\times 10^{-4}$.  This is quite conservative as, at LSND, these decays in flight occur in a vacuum region following the target, but we require no such vacuum region.  In this study we have assumed an uncertainty in this ratio of only 5\% (of $5\times 10^{-4}$), however in \cite{adsprelim} we found that the results are not significantly changed using an uncertainty of 25\%.  We have found that this background is significant only for baselines of less than 5 km.

\subsection{Atmospheric Electron Neutrinos}

There is an irreducible background due to low energy atmospheric neutrinos.  Atmospheric electron antineutrinos in the energy range of 30 MeV to 53 MeV IBD capture on hydrogen identically to the signal $\overline{\nu}_e$, although the shape of the background is quite different from that of the signal and so a shape analysis can be applied.  Atmospheric $\nu_e$ and $\overline{\nu}_e$  of higher energy may enjoy quasielastic (QE) charged current (CC) interactions with oxygen in the water Cherenkov detector.   These QE background events often result in the creation of additional particles which can be used to veto them.

We use the unoscillated $\overline{\nu}$ and $\nu$ spectra at the Kamioka mines given in Ref. \cite{honda}  at energies above 100 MeV and at lower energies we use the spectra available on M.~Honda's website \cite{hondasito}.  % We extrapolate these spectra down to our energies of interest by fitting to the shape of atmospheric neutrinos expected at Kamioka in Ref. \cite{leebludman}, while using Ref. \cite{honda} to fix the normalization at 100 MeV in each angular bin.  
Neutrino oscillation is performed for several sample points in each angular bin and then angular integration yields the oscillated flux expected at the Kamioka site.  

IBD events are charged current interactions of $\overline\nu_e$ with a free proton which yield a neutron and a positron.  We will approximate free protons to be 11\% of the fiducial mass, and so 2.4~kton and 62~kton at SK and HK respectively.  Therefore there will be $1.4\times 10^{33}$ free protons in SK and $3.7\times 10^{34}$ in HK.  Multiplying the oscillated neutrino flux, the total detector cross section and the 6-year runtime we find 9 IBD background events between 30 and 54 MeV at SK and 225 at HK.   

We use GENIE simulations \cite{genie} to calculate the number of electron and positron events that will result from charged current quasielastic interactions (CCQE) on oxygen.  In contrast with IBD $\overline{\nu}_e$, the corresponding $\nu_e$ and $\overline{\nu}_e$ energies are generally much higher than 50 MeV, however due to the Fermi momentum of the nucleon target, the resulting charged lepton energy can be in our signal range.  We fold the resulting electron and positron spectra with the oscillated atmospheric $\nu_e$ and $\overline{\nu}_e$ spectra to derive the expected atmospheric $\nu_e$ and $\overline{\nu}_e$ backgrounds.   We consider the sum of the IBD  $\overline{\nu}_e$ and CCQE $\nu_e$ and $\overline{\nu}_e$ backgrounds.  % and normalize it to the best fit to SK's observations in Ref.~\cite{sksn}.   The resulting background is the lowest curve in each panel of Fig.~\ref{eventifig}.  

SK-IV is able to detect $\gamma$ from the H capture of the neutron arising from IBD interactions of $\overline{\nu}$ \cite{skwendell} with an efficiency which is now about 20\%.  We find that a double coincidence requirement could in principle eliminate most of the CCQE background.  However, to be conservative we have not applied this veto in our analysis.

%({\bf{Write the results when they are available}})

%Due to the stronger tangential magnetic field, the atmospheric neutrino flux at JUNO will be lower than at Kamioka and RENO-50, and so in the case of $\mu$DARTS at JUNO this procedure will overestimate the atmospheric neutrino background.

\subsection{Invisible Muons}

TNT2K faces an additional, larger, background.  Charged current interactions of atmospheric muon neutrinos on oxygen in the detector will produce muons.   Those muons below the Cherenkov threshold will be invisible to SK and HK and those that decay will produce electrons or positrons whose signal constitutes the invisible muon background.  More specifically, $\nu_{\mu}$ yield $\mu^-$ of which about $20\%$ will be absorbed by the oxygen in the water and two thirds of the relevant events will produce no neutrons and so will not yield a false double coincidence.   On the other hand $\overline{\nu}_\mu$ produce $\mu^+$, which are not absorbed, and usually yield neutrons and so a fake double coincidence.  While this makes each $\overline{\nu}_\mu$ event more dangerous, the higher cross-section for $\nu_\mu$ events in this energy range in fact implies that they provide the dominant background.  While most of these $\nu_\mu$ events can in principle be vetoed by requiring a double coincidence with neutron capture, again we have chosen not to apply this veto in our full analysis.

The shape of the background is the well known Michel spectrum, as it results from the decay of muons into electrons.  It is identical to our $\mu^+$DAR $\overline{\nu}$ spectrum.  However, the signal $\overline{\nu}_e$ are detected via IBD which creates $e^+$ of energy 1.3 MeV lower than the original antineutrino, and so in fact the signal spectrum is 1.3 MeV lower than the invisible muon background spectrum.  Nonetheless, given the energy resolutions of SK and HK, this shift is of limited use in distinguishing the signal and background.  The normalization of the invisible muon background has been well-measured by SK \cite{sksn} although, as these neutrinos are ultimately generated by cosmogenic muons, the rate may be time-dependent at the 10-20\% level.   In summary, in combination with some accelerator down-time, not only the shape but also the invisible background muon rate will be known rather precisely.

%Note that if the cyclotron is off for 20\% of the time during 6 years of live time, HK may expect 2138 invisible muon events during the dead time, of which 575 (1050) would pass the vetoes with (without) Gd.  The error on the background calculations may well exceed 30\%, but this beam off measurement will allow a background measurement with a precision of 4\% (3\%).  As the shape of the invisible background is known and its normalization will be known precisely, this background can be subtracted reliably.

In our calculations, we have assumed that the normalization of the invisible muon background and $\nu_e$ backgrounds are known with 7\% and 10\% precisions respectively, reflecting statistical fluctuations in the SK sample \cite{sksn} and, since SK measured the total background, we assume an error correlation of $-0.24$ between the two backgrounds.

The CC events that produce the muon also produce other particles.  In fact in the case of $\Delta$ resonance CC events, which account for most of the events in the second column of Table~\ref{invtab}, additional pions and generally other particles are always produced, such as $\gamma$ \cite{pregadzooks}.  These other particles in general produce some effect which is visible at the detector, allowing for vetoes of the background.  We have studied the potential veto efficiencies.  In our main analysis we use the selection of vetoes summarized in Table.~\ref{vetotab}.  We make the crude approximation that all $\gamma$ events, except for those resulting from neutron capture, can be identified.  

For example, consider the following event.  A $\Delta$ and an invisible $\mu$ are created in a CC event.  The $\mu$ decays after a few microseconds and produces an electron, leading to an electron-like ring.  Let us call this ring number one.  The $\Delta$ decays, for example, into a charged $\pi$.  The $\pi$ decays quickly, yielding a $\mu$ which decays after a few microseconds yielding another $e$ and so a second electron-like ring, let us call it ring number 2.  Note that in this case, as the two electrons are created from the decays of {\it{distinct}} muons, which are created essentially simultaneously, the expected time difference between rings number one and two is roughly the muon lifetime.   This background can essentially be eliminated by removing events with two electron-like rings.  We veto events with multiple electron-like rings.% and also events in which multiple rings appear separated by a few microseconds.

%even if $\pi^+$ are below the Cherenkov threshold they will decay to $\mu^+$ which will decay to $e^+$ and so instead of one electron-like event there will be two.    Due to the muon lifetime, these extra events will often be separated from the invisible muon's electron event by several microseconds.  However, by vetoing all events for which there are other events within a few microseconds, the invisible muon background can be reduced considerably \cite{gadzooks}, essentially eliminating all events from neutrinos with energies above 300 MeV.

\begin{table}%[position specifier]
\centering
%\begin{tabular}{c|l|l|l}
\begin{tabular}{|c|}
\hline
Proposed Selection Cuts\\
\hline\hline
%40 km&&\%\ (\%)&\%\ (\%)\\
%\hline
All events that occur during the J-PARC beam spill\\
\hline
Events with multiple rings\\
\hline
Events with $\nu$ energies outside the 30-55 MeV window\\
\hline
Events which are not fully contained\\
\hline
Events within 1 ms of electron or muon rings\\
\hline
Events followed by $\gamma$ emission within 1 ms\\
\hline
%with 1 $n$&$7/11$&$6/10$&$1/1$\\
%\hline
%no $\gamma$&$17/8$&$15/7$&$2/0$\\
%\hline
%with 1 $n$, no $\gamma$&$4/6$&$3/6$&$1/0$\\
%\hline
\end{tabular}
\caption{Proposed background selection cuts}
\label{vetotab}
\end{table}

We have folded the results of GENIE simulations of atmospheric neutrino events in water with the atmospheric neutrino fluxes of Ref. \cite{honda}, oscillated using the neutrino mass matrix parameters of Subsec. \ref{angoli}, to determine the veto efficiencies.  On the other hand, we use Super-K measurements to fix the overall normalization of the invisible muon background which we use in this section.  GENIE yielded the number of invisible $\mu^+$ and $\mu^-$.  Essentially all of the $\mu^+$ and 80\% of the $\mu^-$ come to rest and then decay in water, and so the number of background events is the number of $\mu^+$ plus 80\% of the number of $\mu^-$ events.   

We find that few invisible muon events arise from NC interactions, most of which produce additional particles which can be used to veto them.  Therefore in our main analysis we only consider the CC invisible muon background.  In Table \ref{invtab} we provide the number of invisible $\mu$ events caused by interactions of $\nu$ and $\overline\nu$ per six years.   The first row includes all events.  The second row provides the number of events in which, according to our GENIE simulations, no $\gamma$  is emitted as the struck nucleus de-excites.  Note that $\gamma$ may also be emitted by neutron capture on H, however the energy of this $\gamma$ is lower than that which accompanies a nuclear de-excitation in most of our simulations.  Furthermore we suspect that the $\gamma$ emission from nuclear de-excitation is generally much faster than $\gamma$ capture on H and so these may be distinguished.  If a detector is loaded with $0.1\%$ Gd then the vast majority of $n$ captures will be on Gd.  This capture results in the production of 3-5 $\gamma$ which have similar energy distributions to de-excitation $\gamma$, however the $n$ capture is delayed by 10s of $\mu$sec and this delay can be used to discriminate the two kinds of $\gamma$.  

In our analysis we use the second row, corresponding to a veto of events with additional $\gamma$.  We restrict our attention to neutrinos with $E_\nu\leq$ 300 MeV because we find that higher energy $\nu$ essentially always create extra rings which can be used for a veto.  According to the second row of the first column of Table~\ref{invtab}, this leaves us with 149 invisible $\mu^-$ events and 73 invisible $\mu^+$ events, for a total of 222 events.  Furthermore, to reduce the spallation, diffuse supernova backgrounds and in particular quasi-elastic interactions, we restrict our attention to events in which the final $e$ energy yields a reconstructed $\nu$ energy of 30-50 MeV.  This latter condition leaves 155 of the 222 original invisible $\mu$ events, as can be seen in Fig.~\ref{eventifig}.  

In the third row no condition is placed on $\gamma$ but we consider only events which yield precisely 1 $n$.  As we do not consider Gd doping, the efficiency with which such a veto may be implemented will be limited and so it is not considered in our analysis.  Finally in the last row we impose both the single $n$ and the no de-excitation $\gamma$ requirements.

%with and without a veto of events with $\gamma$ and also with and without a single neutron double coincidence requirement.   Vetoing events with $\gamma$'s, for a 6 year run, at SK we expect 53 (134)  invisible $\mu$ events originated from charged current interactions of neutrinos with energy under 300 MeV  with (without) a double coincidence requirement.  We find that most invisible muon events arise from NC interactions, most of which produce additional particles which can be used to veto them.  In the determination of $\delta$ we do not impose a double coincidence requirement, but we consider only events between 30 and 55 MeV which, as reported in Fig.~\ref{eventifig}, in 6 years leaves 96 of the 134 invisible $\mu$ events. % This can be compared with the expected signal of 350 events at SK and about 4500 at HK if $\delta=0$.

%Note that in the rest of this note we do not use the invisible muon background normalizations summarized in Table~\ref{invtab}, but rather we normalize the background to that observed at SK in Ref.~\cite{sksn}, yielding the backgrounds reported in Fig.~\ref{eventifig}.  

As SK had a hard trigger during the SK runs used in Ref.~\cite{sksn},  it could not yet detect many of the low energy $\gamma$'s used in the various cuts in Table.~\ref{invtab}.  Thus it is not surprising that the background rate observed at SK is between the total and no $\gamma$ rates in Table.~\ref{invtab}.% that the background rate that we find with no double coincidence requirement is similar to that observed by SK.  The SK rate is however a bit lower, perhaps as a result of vetoes resulting from additional particles besides $\gamma$.  

\begin{table}%[position specifier]
\centering
\begin{tabular}{c|l|l|l|l}
&%total&
CC: $E_\nu\leq$300&CC: $E_\nu\geq$300&NC: $E_\nu\leq$300&NC: $E_\nu\geq$300\\
\hline\hline
%40 km&&\%\ (\%)&\%\ (\%)\\
%\hline
all inv. $\mu$&
$343/135$&$83/12$&$0/0$&$38/20$\\
\hline
no $\gamma$&%$102/45$&
$149/73$&$20/2$&$0/0$&$17/8$\\
\hline
1 $n$&%$41/64$&
$54/99$&$13/5$&$0/0$&$21/12$\\
\hline
1 $n$, no $\gamma$&%$21/36$&
$30/58$&$5/2$&$0/0$&$10/5$\\
%all inv. $\mu$&%$258/89$&
%$208/82$&$50/7$&$0/0$&$23/12$\\
%\hline
%no $\gamma$&%$102/45$&
%$90/44$&$12/1$&$0/0$&$10/5$\\
%\hline
%1 $n$&%$41/64$&
%$33/60$&$8/3$&$0/0$&$13/7$\\
%\hline
%1 $n$, no $\gamma$&%$21/36$&
%$18/35$&$3/1$&$0/0$&$6/3$\\
\hline
\end{tabular}
\caption{Number of $\nu_\mu/\overline\nu_\mu$ invisible $\mu$ events per 6 years expected at SK including various veto conditions.  The columns from left to right correspond to CC events with neutrino energies less than and greater than 300 MeV and NC events with neutrino energies less than and greater than 300 MeV.}
\label{invtab}
\end{table}

\begin{table}%[position specifier]
\centering
\begin{tabular}{c|l|l}
Event Type&SK at 15 km&HK at 23 km\\
\hline\hline
%40 km&&\%\ (\%)&\%\ (\%)\\
%\hline
IBD Signal $\delta=0^\circ$&298&3419\\
\hline
IBD Signal $\delta=90^\circ$&325&4549\\
\hline
IBD Signal $\delta=180^\circ$&240&3556\\
\hline
IBD Signal $\delta=270^\circ$&214&2426\\
\hline
Invisible $\mu$ Background&155&3862\\
\hline
Atmos $\nu_e$ Background&26&639\\
\hline
$\mu^-$DAR Background&2.7&27\\
\hline
CCQE $\nu_e-{}^{16}$O Background&32&332\\
\hline
$\nu_e-e^-$ Elastic Background&46&478\\
\hline
\end{tabular}
\caption{Total number of signal and background events expected in 6 years, implementing the vetoes in Table~\ref{vetotab}.}
\label{eventitottab}
\end{table}

\section{Sensitivity to $\delta$}

\subsection{Optimizing the Baselines}

Where should the $\mu$DAR source be placed?  As the event rate will be lower at a far detector, it is reasonable that the far detector be larger.   This suggests that the source be to the north of SK.  Furthermore, a maximum synergy is achieved when the baselines to SK and HK are as different as possible.  The maximum difference is the distance between the Mozumi and Tochibora mines, 8 km.  Therefore we will always assume that the baseline to HK is 8 km larger than that to SK, and will optimize the baseline to SK.

We consider three different cases.  In each case a 6 year $\mu$DAR run is combined with 12 years of T2K (and 6 years of T2HK) in $\nu$ mode only, as we find that the time in $\overline{\nu}$ mode reduces the performance of TNT2K.  In the first case, only SK is considered.  In the second only 20\% of HK will be in operation for 6 years.  Recall that HK consists of 10 identical 100 kton modules, each with a 56 kton fiducial volume.  Therefore, 20\% of HK corresponds to the construction of two modules.  We refer to this case as HK/5.  In the third, the full HK is assumed to be in operation.  In each case we assume that the neutrino mass hierarchy, NH or IH, is known but we consider both hierarchies.

\begin{figure} %[!tph]
\begin{center}
\includegraphics[angle=-0,width=6.2in]{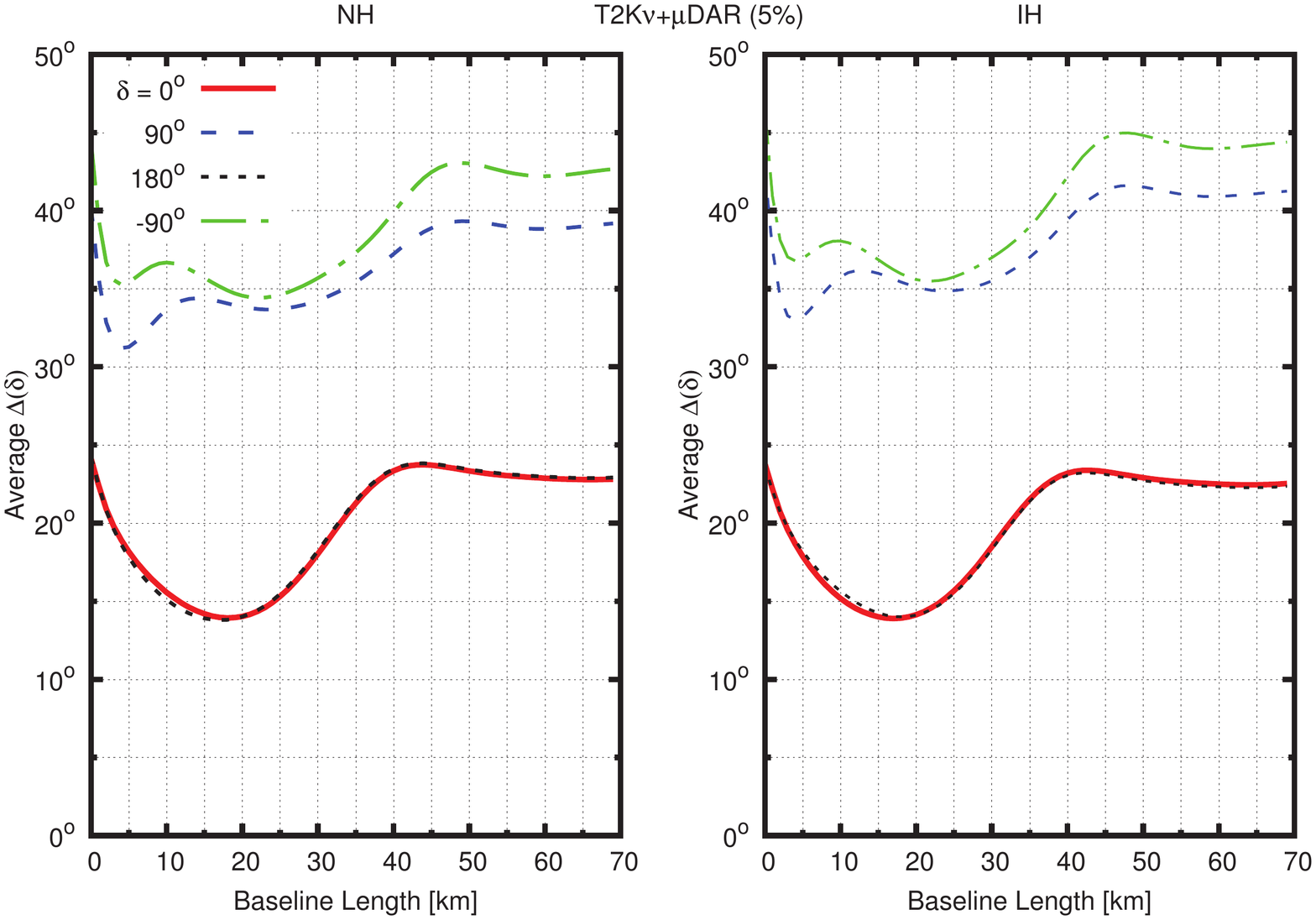}
\caption{The precision with which $\delta$ can be determined using SK only as a function of the $\mu$DAR baseline.  The precision quoted is the average of the upper and lower uncertainties.  Both hierarchies are considered, but it is assumed that the hierarchy is known.}
\label{skbasefig}
\end{center}
\end{figure}

We calculate, for each baseline to SK, for $\delta_{\rm{true}}=0^\circ,\ 90^\circ,\ 180^\circ$ and $270^\circ$, the fit values of $\delta_{\rm{fit}}$ for which the best fit of one theoretical data set to the other yields $\chi^2=1$, corresponding to an expected $\Delta\chi^2=1$ for a fit of real data.  As expected this occurs for one value of $\delta_{\rm{fit}}$ which is larger than $\delta_{\rm{true}}$ and one which is smaller.  These intervals are roughly symmetric about the true value of $\delta$.  In Fig.~\ref{skbasefig} we report half of the size of the interval, which is approximately the 1$\sigma$ precision which can be expected in a measurement of $\delta$ at TNT2K with SK only.  

One sees that the optimal baseline depends on sin$(\delta)$.  For sin$(\delta)=0$ it is 15-20 km, where the expected uncertainty on $\delta$ is about $14^\circ$ whereas for maximal CP violation it is 20-30 km, where the expected uncertainty is about $34^\circ$.  If 15 km is adopted then one finds that the uncertainty lies in the range $14^\circ-36^\circ$, not far from the optimal for any value of $\delta$.  However we have also found that, while 15 km yields a competitive measurement of $\delta$, a longer baseline would yield a more robust breaking of the $\delta\rightarrow 180^\circ-\delta$ degeneracy.

\begin{figure} %[!tph]
\begin{center}
\includegraphics[angle=-0,height=3.9in,width=6.2in]{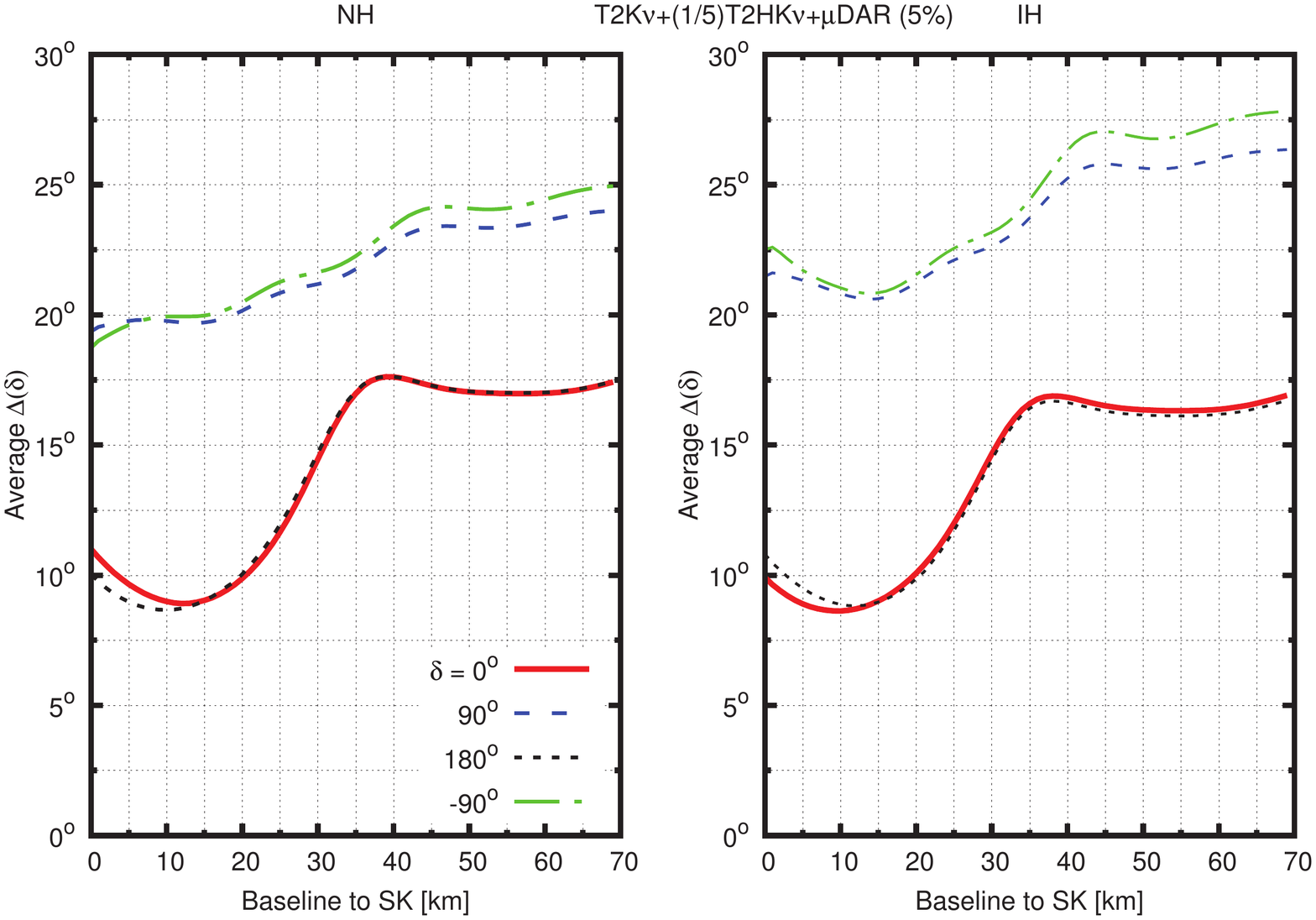}
\includegraphics[angle=-0,height=3.9in,width=6.2in]{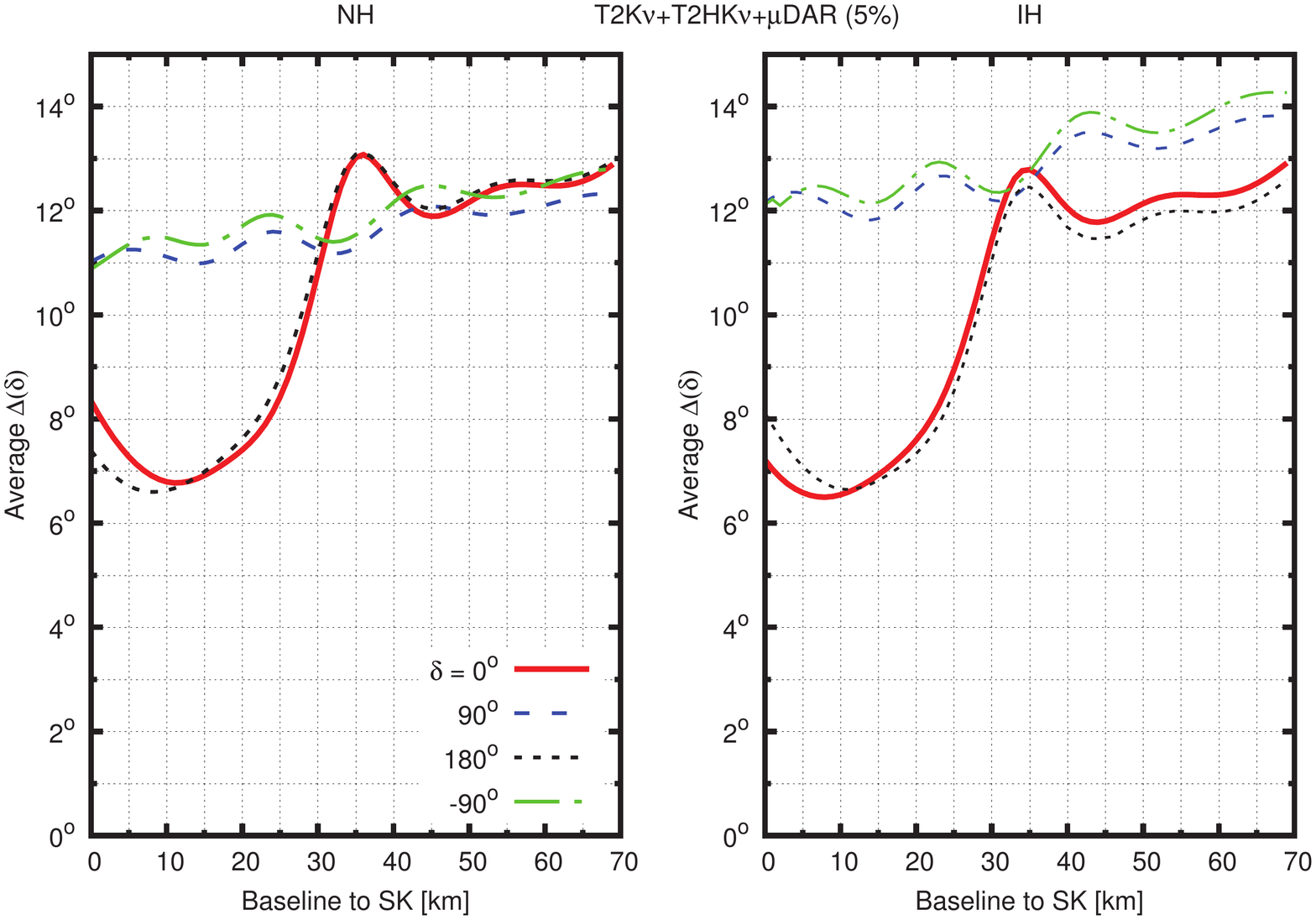}
\caption{The precision with which $\delta$ can be determined using SK and one fifth of HK (top) or all of HK (bottom) as a function of the $\mu$DAR baseline to SK, as in Fig.~\ref{skbasefig}.  The baseline to HK is 8 km greater.}
\label{hkbasefig}
\end{center}
\end{figure}

In Fig.~\ref{hkbasefig} we present the precision with which $\delta$ can be measured in the cases with one fifth of HK (the upper panels) and all of HK (the lower panels) , for a given baseline from the $\mu$DAR source to SK.  Not surprisingly, as HK is 8 km further, the optimal baseline to SK is now shorter than the SK only case.   In general the most precise determination of $\delta$ occurs for a baseline of roughly 15 km to SK and 23 km to HK.  With one fifth of HK (all of HK) the precision with which $\delta$ can be measured ranges from $9^\circ$ ($7^\circ$) for no CP violation to $20^\circ$ ($11^\circ$) in the case of maximal CP violation.   

This can be compared with the performance of T2HK without $\mu$DAR and with the J-PARC beam running for 1.5 years in $\nu$ mode and 4.5 years in $\overline{\nu}$ mode, with the full HK.  In that case one expects to measure $\delta$ \cite{HK2} with a precision of $9^\circ-24^\circ$:  {\it{With $\mu$DAR and one fifth of HK, one can determine $\delta$ more precisely than with all of HK and no $\mu$DAR.}}

\subsection{Measuring $\delta$ with TNT2K}

\begin{figure} %[!tph]
\begin{center}
\includegraphics[angle=-0,height=3.9in,width=5.1in]{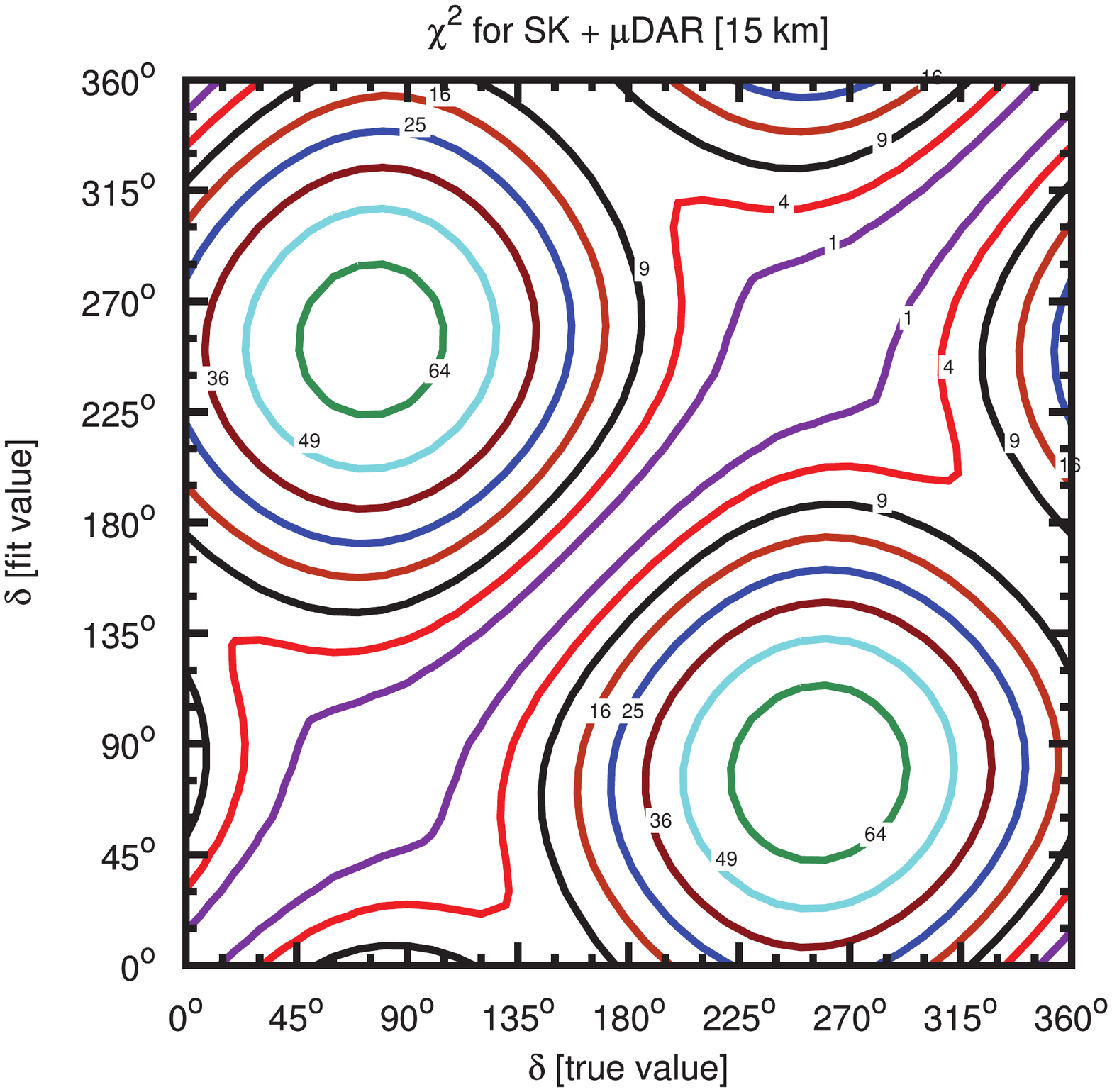}
\includegraphics[angle=-0,height=3.9in,width=5.2in]{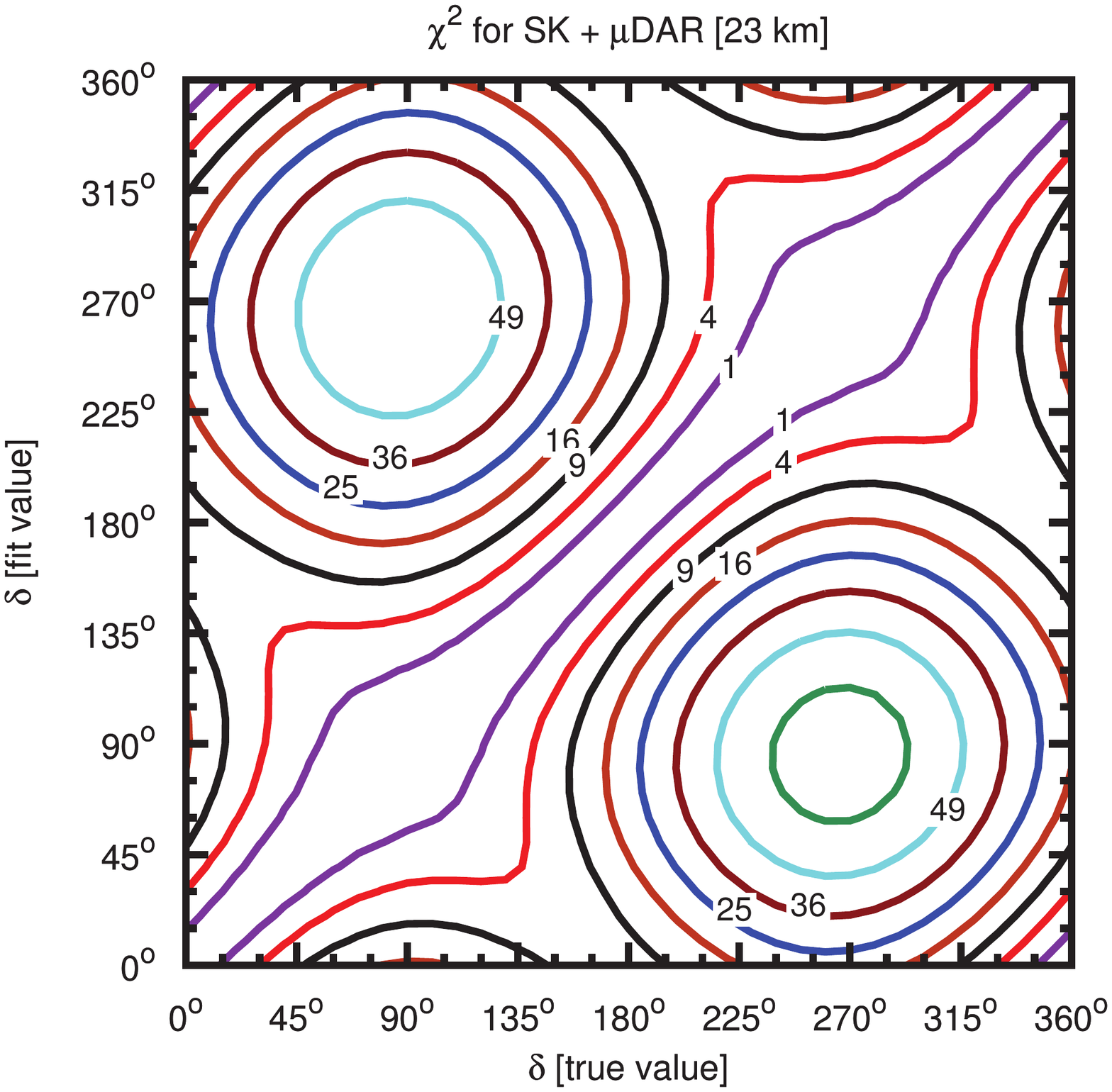}
\caption{$\chi^2$ value of each trial value of  $\delta$ vs each true value assuming the normal hierarchy, using SK only at 15 km (top) and 23 km (bottom) with 6 years of $\mu$DAR and 12 years of T2K operating in $\nu$ mode.  From the $x$-axis one observes that maximal and null CP-violation can be distinguished at about $4-5\sigma$ while, with null CP-violation, the sign of cos$(\delta)$ can be determined with $2-3\sigma$ of confidence.  Null CP violation can be excluded at more than 3$\sigma$ of confidence for more than half of the values of $\delta$.}
\label{chit2kfig}
\end{center}
\end{figure}

In Fig.~\ref{chit2kfig} we plot the $\chi^2$ value of the best fit of the $\delta_{\rm{fit}}$ theoretical spectrum to the theoretical spectra of $\delta_{\rm{true}}$ for SK only, at 15km and also at 23 km assuming the normal hierarchy.  Looking at the $x$-axis for the value of $\delta_{\rm{true}}$, one observes that for a 23 km baseline maximal CP-violation, corresponding to $\delta=90^\circ$ or $270^\circ$ can be distinguished from no CP-violation, corresponding to $\delta=0^\circ$ and indeed also to $180^\circ$, at about $4\sigma$ of confidence and in fact nearly $5\sigma$ for $\delta=270^\circ$.    With a 15 km baseline, $\delta=0^\circ$ can be distinguished from $\delta=90^\circ$ and $270^\circ$ with a bit under $4\sigma$ and $6\sigma$ respectively.   In both cases, $\delta=0^\circ$ and $180^\circ$ can only be distinguished at $2-3\sigma$ of confidence.

\begin{figure} %[!tph]
\begin{center}
\includegraphics[angle=-0,height=3.9in,width=5.2in]{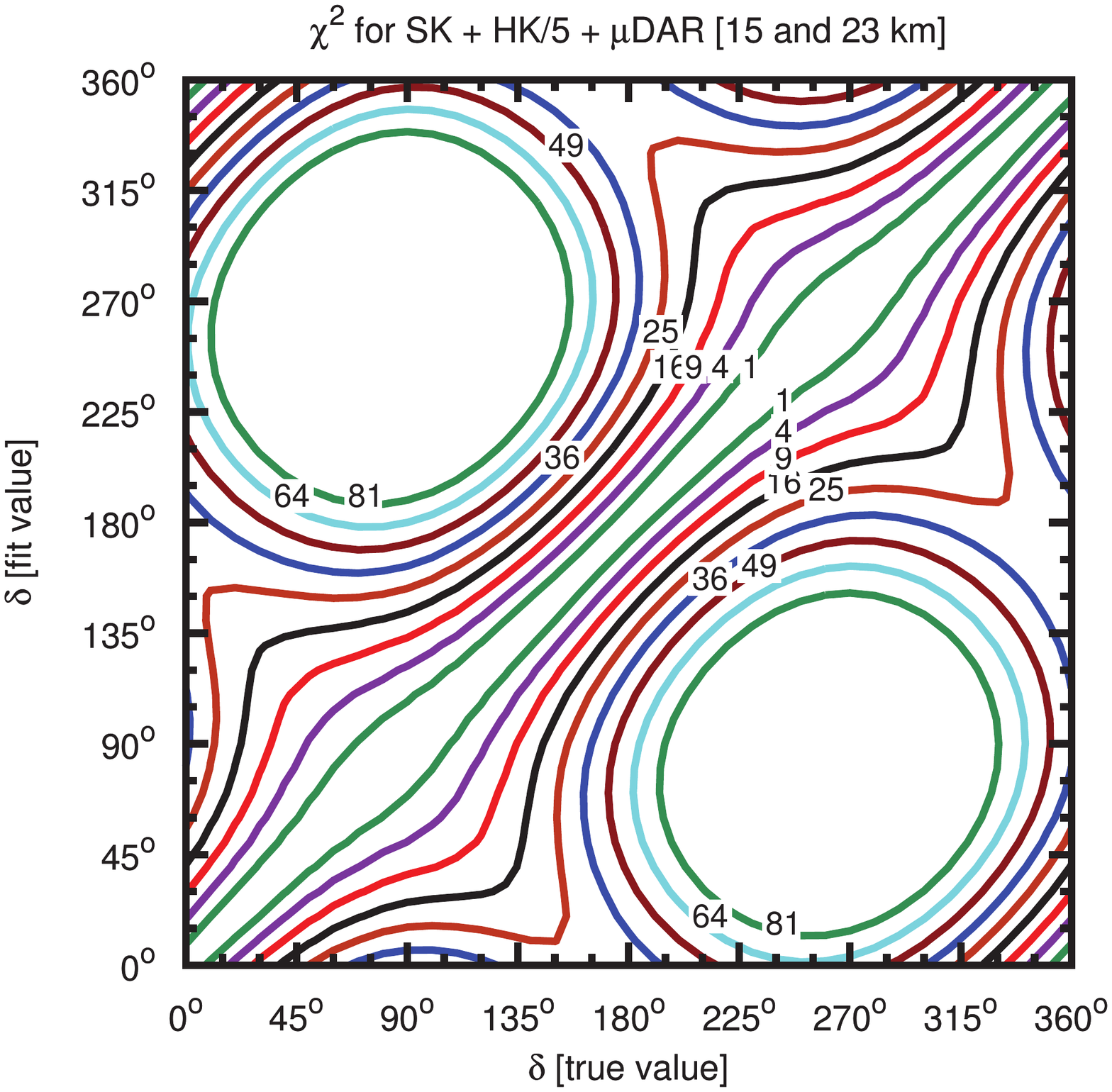}
\includegraphics[angle=-0,height=3.9in,width=5.2in]{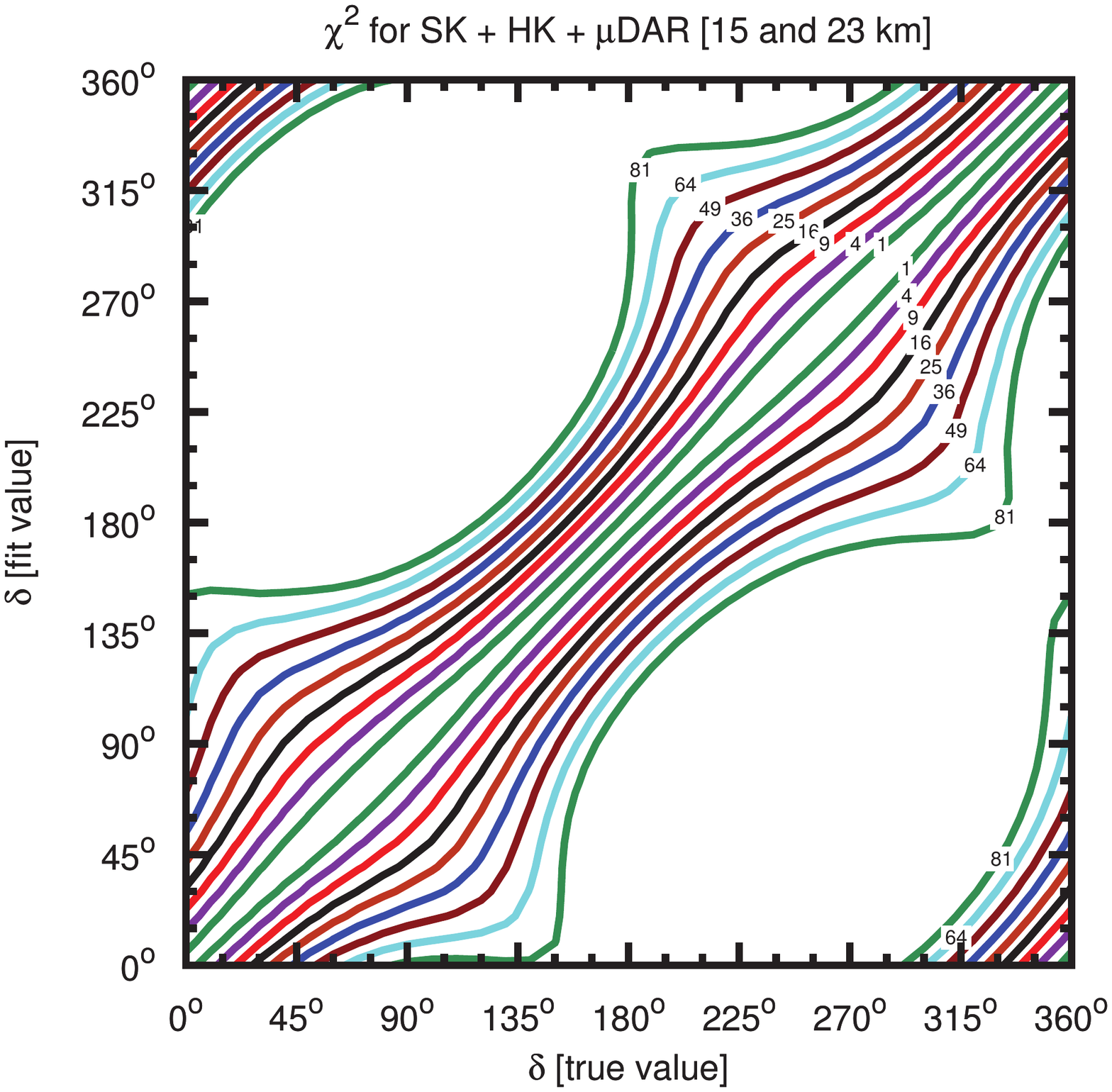}
\caption{As in Fig. \ref{chit2kfig} but now including HK (bottom) and also one fifth of HK (top).  With one fifth of HK, maximal and null CP-violation can be distinguished at more than $7\sigma$ while, with null CP-violation, the sign of cos$(\delta)$ can be determined with nearly $6\sigma$ of confidence.  Null CP violation can be excluded at more than 5$\sigma$ of confidence for more than half of the values of $\delta$.}
\label{errorit2hkfig}
\end{center}
\end{figure}

In Fig.~\ref{errorit2hkfig} we consider SK at 15 km and one fifth or all of HK at 23 km.  With just one fifth of HK, one sees that maximal CP violation, $\delta=90^\circ$ ($270^\circ$) can be distinguished from $\delta=0$ at more than $6\sigma$ ($7\sigma$).  Also $\delta=0^\circ$ and $\delta=180^\circ$ can be distinguished with nearly $6\sigma$ of confidence.  Thus, the $\delta\rightarrow 180^\circ-\delta$ degeneracy, for large CP violation, is completely broken already with one fifth of HK.  On the other hand, with the full HK, these distinctions can be made at more than $9\sigma$.  Thus in general the full HK only serves to provide a precise determination of $\delta$, one fifth of HK is quite sufficient to qualitatively understand leptonic CP violation.

\section{Conclusions}

A 7 MW proton accelerator, with a proton energy of 600 MeV-1.5 GeV, when striking a medium to high $Z$ target creates 30-50 MeV $\overline{\nu}_\mu$ via $\mu^+$ decay at rest.  We have advocated placing such an accelerator 15 km north of SK.  The oscillated $\overline{\nu}_e$ can be detected by SK via IBD.  The $\overline{\nu}$ spectrum is known quite precisely as is the IBD cross-section, thus systematic errors are small in the determination of the CP-violating phase $\delta$.  Together with 12 years of T2K, this setup allows $\delta$ to be measured with a precision of $14^\circ-36^\circ$ in 6 years.  Maximal and null CP violation can be distinguished at about $4-5\sigma$ and $\delta=0^\circ$ can be distinguished from $\delta=180^\circ$ with $2-3\sigma$ of confidence, a very difficult task for conventional beam experiments.

If just one fifth of HK is built, corresponding to two of the ten planned modules, then at the preferred Tochibora mine it will be 23 km from the $\mu$DAR source.  This will allow an excellent determination of $\delta$, with a precision of $9^\circ-20^\circ$ in 6 years.  Maximal and minimal CP violation can be distinguished with $6-8\sigma$ of confidence and the sign of cos($\delta$) at well over $5\sigma$.

The required accelerator is not far beyond the state of the art, and as the neutrinos are created from decay at rest the only requirement is that the $\pi$ and $\mu$ stop in the target.  This means that many other physics programs can be done simultaneously with the same accelerator, for example it can run an accelerator driven subcritical nuclear reactor.  Such accelerators in fact have just the specifications that we require.

\section* {Acknowledgement}
\noindent
We are privileged to thank Emilio Ciuffoli and Xinmin Zhang for useful discussions and correspondence.  We thank Mark Vagins for discussions about GADZOOKS, Prof. Hayato for discussions on supernova relic neutrino measurement at SK, and Costas Andreopoulos for discussions regarding the applicability of GENIE.  JE is supported by NSFC grant 11375201 and a KEK Visiting Fellowship.   The Japan Society for the Promotion of Science (JSPS) has generously provided SFG a postdoc fellowship to do research at KEK, which is deeply appreciated.  This work is supported in part by Grant-in-Aid for Scientific research (No. 25400287) from JSPS, the William F.~Vilas Trust Estate, and by the U.S.\ Department of Energy under the  contract DE-FG02-95ER40896.

%%%%%%%%%%%%%%%%%%%%%%%%%%%%%%%%


\begin{thebibliography}{99}%\setlength{\itemsep}{-2.3mm}

%%%%%%%%%%%%%%%%%%%%%%%%%%%%%%%%%



\bibitem{skwendell}
  R.~Wendell [Super-Kamiokande Collaboration],
  ``Atmospheric Results from Super-Kamiokande,''
  arXiv:1412.5234 [hep-ex].


\bibitem{HK}
 K.~Abe, T.~Abe, H.~Aihara, Y.~Fukuda, Y.~Hayato, K.~Huang, A.~K.~Ichikawa and M.~Ikeda {\it et al.},
  ``Letter of Intent: The Hyper-Kamiokande Experiment --- Detector Design and Physics Potential,''
  arXiv:1109.3262 [hep-ex].

\bibitem{HK2}
  K.~Abe {\it et al.}  [Hyper-Kamiokande Proto- Collaboration],
  ``Physics Potential of a Long Baseline Neutrino Oscillation Experiment Using J-PARC Neutrino Beam and Hyper-Kamiokande,''
  %Submitted to: PTEP (2015)
  [arXiv:1502.05199 [hep-ex]].


\bibitem{atomic}
T. Zhang, J.~Yang, M.~Li, L.~Xia, S.~An, Z.~Yin, J.~Zhong,
F.~Yang, W.~Joho, A.~Adelmann, P.~Sigg,
``Conceptual design of an 800 MeV high power proton driver,"
NIM B {\bf 269} (2011) 2964.


\bibitem{graphite}
C. Tschal\"ar,
``High-Power Proton Beam Dump with Uniform Graphite Depth,"
Bates Preprint B/IR{}\#{}12‐01.

\bibitem{ads}
  Z.~Li, P.~Cheng, H.~Geng, Z.~Guo, Y.~He, C.~Meng, H.~Ouyang and S.~Pei {\it et al.},
  ``Physics design of an accelerator for an accelerator-driven subcritical system,''
  Phys.\ Rev.\ ST Accel.\ Beams {\bf 16} (2013) 8,  080101.

\bibitem{lsnd97}
C.~Athanassopoulos {\it et al.}  [LSND Collaboration],
  ``Measurements of the reactions C-12 (electron-neutrino, e-) N-12 (g.s.) and C-12 (electron-neutrino, e-) N*-12,''
  Phys.\ Rev.\ C {\bf 55} (1997) 2078
  [nucl-ex/9705001].

\bibitem{lsnd01}
 A.~Aguilar-Arevalo {\it et al.}  [LSND Collaboration],
  ``Evidence for neutrino oscillations from the observation of anti-neutrino(electron) appearance in a anti-neutrino(muon) beam,''
  Phys.\ Rev.\ D {\bf 64} (2001) 112007
  [hep-ex/0104049].


\bibitem{daedwhite}
  C.~Aberle, A.~Adelmann, J.~Alonso, W.~A.~Barletta, R.~Barlow, L.~Bartoszek, A.~Bungau and A.~Calanna {\it et al.},
  ``Whitepaper on the DAEdALUS Program,''
  arXiv:1307.2949.

\bibitem{noidarts}
   E.~Ciuffoli, J.~Evslin and X.~Zhang,
  ``The Leptonic CP Phase from Muon Decay at Rest with Two Detectors,''
  JHEP {\bf 1412} (2014) 051
  [arXiv:1401.3977 [hep-ph]].

\bibitem{noiinterf}
  E.~Ciuffoli, J.~Evslin and X.~Zhang,
  ``Optimizing Medium Baseline Reactor Neutrino Experiments,''
  Phys.\ Rev.\ D {\bf 88} (2013) 033017
  [arXiv:1302.0624 [hep-ph]].

\bibitem{parkedegen}
  H.~Minakata and S.~J.~Parke,
  ``Correlated, Precision Measurements of $\theta_{23}$ and $\delta$ using only the Electron Neutrino Appearance Experiments,''
  Phys.\ Rev.\ D {\bf 87} (2013) 113005
  [arXiv:1303.6178 [hep-ph]].

\bibitem{shaofeng1}
  S.~F.~Ge, D.~A.~Dicus and W.~W.~Repko,
  ``Residual Symmetries for Neutrino Mixing with a Large $\theta_{13}$ and Nearly Maximal $\delta_D$,''
  Phys.\ Rev.\ Lett.\  {\bf 108} (2012) 041801
  [arXiv:1108.0964 [hep-ph]].

\bibitem{shaofeng2}
  S.~F.~Ge, D.~A.~Dicus and W.~W.~Repko,
  ``$\Z_2 $ Symmetry Prediction for the Leptonic Dirac CP Phase,''
  Phys.\ Lett.\ B {\bf 702} (2011) 220
  [arXiv:1104.0602 [hep-ph]].

\bibitem{peter1}
  P.~Ballett, S.~F.~King, C.~Luhn, S.~Pascoli and M.~A.~Schmidt,
  %``Testing atmospheric mixing sum rules at precision neutrino facilities,''
  Phys.\ Rev.\ D {\bf 89} (2014) 1,  016016
  [arXiv:1308.4314 [hep-ph]].

\bibitem{shaofeng3}
  A.~D.~Hanlon, S.~F.~Ge and W.~W.~Repko,
  ``Phenomenological consequences of residual $ \mathbb{Z}^s_2$ and $ \overline {\mathbb{Z}}^s_2$ symmetries,''
  Phys.\ Lett.\ B {\bf 729} (2014) 185
  [arXiv:1308.6522 [hep-ph]].

\bibitem{petcos1}
 S.~T.~Petcov,
  ``Predicting the values of the leptonic CP violation phases in theories with discrete flavour symmetries,''
  Nucl.\ Phys.\ B {\bf 892} (2015) 400
  [arXiv:1405.6006 [hep-ph]].
 
\bibitem{peter2}
  P.~Ballett, S.~F.~King, C.~Luhn, S.~Pascoli and M.~A.~Schmidt,
  ``Testing solar lepton mixing sum rules in neutrino oscillation experiments,''
  JHEP {\bf 1412} (2014) 122
  [arXiv:1410.7573 [hep-ph]].

\bibitem{petcos2}
  I.~Girardi, S.~T.~Petcov and A.~V.~Titov,
  ``Determining the Dirac CP Violation Phase in the Neutrino Mixing Matrix from Sum Rules,''
  arXiv:1410.8056 [hep-ph].

\bibitem{titov}
  I.~Girardi, S.~T.~Petcov and A.~V.~Titov,
  ``Predictions for the Dirac CP Violation Phase in the Neutrino Mixing Matrix,''
  arXiv:1504.02402 [hep-ph].


\bibitem{daed}
J.~Alonso, F.~T.~Avignone, W.~A.~Barletta, R.~Barlow, H.~T.~Baumgartner, A.~Bernstein, E.~Blucher and L.~Bugel {\it et al.},
  ``Expression of Interest for a Novel Search for CP Violation in the Neutrino Sector: DAEdALUS,''
  arXiv:1006.0260 [physics.ins-det].

\bibitem{sksnold}
  M.~Malek {\it et al.}  [Super-Kamiokande Collaboration],
  ``Search for supernova relic neutrinos at SUPER-KAMIOKANDE,''
  Phys.\ Rev.\ Lett.\  {\bf 90} (2003) 061101
  [hep-ex/0209028].

\bibitem{sksn}
  K.~Bays {\it et al.}  [Super-Kamiokande Collaboration],
  ``Supernova Relic Neutrino Search at Super-Kamiokande,''
  Phys.\ Rev.\ D {\bf 85} (2012) 052007
  [arXiv:1111.5031 [hep-ex]].

\bibitem{sksn3}
  H.~Zhang {\it et al.}  [Super-Kamiokande Collaboration],
  ``Supernova Relic Neutrino Search with Neutron Tagging at Super-Kamiokande-IV,''
  Astropart.\ Phys.\  {\bf 60} (2014) 41
  [arXiv:1311.3738 [hep-ex]].

\bibitem{campimag}
${\rm{http://www.ngdc.noaa.gov/geomag/WMM/icons/WMM2010\_ H.png}}$


\bibitem{atm1989}
  G.~Barr, T.~K.~Gaisser and T.~Stanev,
  ``Flux of Atmospheric Neutrinos,''
  Phys.\ Rev.\ D {\bf 39} (1989) 3532.

\bibitem{honda}
  M.~Sajjad Athar, M.~Honda, T.~Kajita, K.~Kasahara and S.~Midorikawa,
  ``Atmospheric neutrino flux at INO, South Pole and Pyh\'asalmi,''
  Phys.\ Lett.\ B {\bf 718} (2013) 1375
  [arXiv:1210.5154 [hep-ph]].


\bibitem{pdg}
 J. Beringer et al. (Particle Data Group), Phys. Rev. D86, 010001 (2012) 

\bibitem{dayaboston}
C. Zhang, ``Recent Results From Daya Bay" talk given at Neutrino 2014 on June 3, 2014, Boston University.

\bibitem{t2kdisp2014}
 K.~Abe {\it et al.}  [T2K Collaboration],
  ``Precise Measurement of the Neutrino Mixing Parameter $\theta_{23}$ from Muon Neutrino Disappearance in an Off-Axis Beam,''
  Phys.\ Rev.\ Lett.\  {\bf 112} (2014) 18,  181801
  [arXiv:1403.1532 [hep-ex]].


\bibitem{parke2005}
 H.~Nunokawa, S.~J.~Parke and R.~Zukanovich Funchal,
  ``Another possible way to determine the neutrino mass hierarchy,''
  Phys.\ Rev.\ D {\bf 72} (2005) 013009
  [hep-ph/0503283].


\bibitem{minosmassa}
 P.~Adamson {\it et al.}  [MINOS Collaboration],
  ``Combined analysis of $\nu_{\mu}$ disappearance and $\nu_{\mu} \rightarrow \nu_{e}$ appearance in MINOS using accelerator and atmospheric neutrinos,''
  Phys.\ Rev.\ Lett.\  {\bf 112} (2014) 191801
   arXiv:1403.0867 [hep-ex].

\bibitem{dayam}
  D.~V.~Naumov [Daya Bay Collaboration],
  ``Recent results from Daya Bay experiment,''
  arXiv:1412.7806 [hep-ex].


\bibitem{weit2k}
  S.~K.~Agarwalla, S.~Prakash and W.~Wang,
  ``High-precision measurement of atmospheric mass-squared splitting with T2K and NOvA,''
  arXiv:1312.1477 [hep-ph].

\bibitem{shaofengprec}
  S.~F.~Ge, K.~Hagiwara, N.~Okamura and Y.~Takaesu,
  ``Determination of mass hierarchy with medium baseline reactor neutrino experiments,''
  JHEP {\bf 1305} (2013) 131
  [arXiv:1210.8141 [hep-ph]].

 
\bibitem{t2kexcess}
 K.~Abe {\it et al.}  [T2K Collaboration],
  %``Observation of Electron Neutrino Appearance in a Muon Neutrino Beam,''
  Phys.\ Rev.\ Lett.\  {\bf 112} (2014) 061802
  [arXiv:1311.4750 [hep-ex]].


\bibitem{minoscp}
  P.~Adamson {\it et al.}  [MINOS Collaboration],
  ``Electron neutrino and antineutrino appearance in the full MINOS data sample,''
  Phys.\ Rev.\ Lett.\  {\bf 110} (2013) 17,  171801
  [arXiv:1301.4581 [hep-ex]].


\bibitem{daedseminario}
 J.~Alonso,
  ``High Current H${}_2^+$ Cyclotrons for Neutrino Physics: The IsoDAR and DAE$\delta$ALUS Projects," AIP Conference 
Proceedings {\bf 1525} (2013)  480  [arXiv:1210.3679 [physics.acc-ph]].. 

\bibitem{5anni}
http://www.kek.jp/ja/About/OrganizationOverview/Assessment/Roadmap/roadmap2013-E.pdf

 
\bibitem{gadzooks}
J.~F.~Beacom and M.~R.~Vagins,
  ``GADZOOKS! Anti-neutrino spectroscopy with large water Cherenkov detectors,''
  Phys.\ Rev.\ Lett.\  {\bf 93} (2004) 171101
  [hep-ph/0309300].
  

\bibitem{SK2005}
 Y.~Ashie {\it et al.}  [Super-Kamiokande Collaboration],
  ``A Measurement of atmospheric neutrino oscillation parameters by SUPER-KAMIOKANDE I,''
  Phys.\ Rev.\ D {\bf 71} (2005) 112005
  [hep-ex/0501064].


\bibitem{kaorusk}
 K.~Hagiwara and N.~Okamura,
  ``Re-evaluation of the T2KK physics potential with simulations including backgrounds,''
  JHEP {\bf 0907} (2009) 031
  [arXiv:0901.1517 [hep-ph]].



\bibitem{nupro}
S.-F. Ge,  NuPro: a simulation package for neutrino physics, http://nupro.hepforge.org 

\bibitem{emilio}
This code was written and these checks were performed by E. Ciuffoli.

%\bibitem{globes}
%  P.~Huber, M.~Lindner and W.~Winter,
 % ``Simulation of long-baseline neutrino oscillation experiments with GLoBES (General Long Baseline Experiment Simulator),''
 % Comput.\ Phys.\ Commun.\  {\bf 167} (2005) 195
 % [hep-ph/0407333].
 %P.~Huber, J.~Kopp, M.~Lindner, M.~Rolinec and W.~Winter,
 % ``New features in the simulation of neutrino oscillation experiments with GLoBES 3.0: General Long Baseline Experiment Simulator,''
 % Comput.\ Phys.\ Commun.\  {\bf 177} (2007) 432
 % [hep-ph/0701187].

\bibitem{globest2k} 
  M.~Fechner,
  ``D\'etermination des performances attendues sur la recherche 
  de l'oscillation $\nu_\mu\rightarrow\nu_e$ dans l'experi\'ence T2K depuis 
  l'\'etude des donn\'ees recueilles dans l'\'experience K2K,''
  DAPNIA-06-01-T.
I.~Kato [T2K Collaboration],
  ``Status of the T2K experiment,''
  J.\ Phys.\ Conf.\ Ser.\  {\bf 136} (2008) 022018.
 J.~-E.~Campagne, M.~Maltoni, M.~Mezzetto and T.~Schwetz,
  ``Physics potential of the CERN-MEMPHYS neutrino oscillation project,''
  JHEP {\bf 0704} (2007) 003
  [hep-ph/0603172].



\bibitem{sterile}
  M.~Harada, S.~Hasegawa, Y.~Kasugai, S.~Meigo, K.~Sakai, S.~Sakamoto, K.~Suzuya and E.~Iwai {\it et al.},
  ``Proposal: A Search for Sterile Neutrino at J-PARC Materials and Life Science Experimental Facility,''
  arXiv:1310.1437 [physics.ins-det].
 
\bibitem{sterile2}
  M.~Harada {\it et al.},
  ``Status Report (BKG measurement): A Search for Sterile Neutrino at J-PARC MLF,''
  arXiv:1502.02255 [physics.ins-det].

\bibitem{volpe}
  R.~Lazauskas and C.~Volpe,
  ``Neutrino beams as a probe of the nuclear isospin and spin-isospin excitations,''
  Nucl.\ Phys.\ A {\bf 792} (2007) 219
 % doi:10.1016/j.nuclphysa.2007.06.005
  [arXiv:0704.2724 [nucl-th]].

\bibitem{muonirev}
  D.~F.~Measday,
  ``The nuclear physics of muon capture,''
  Phys.\ Rept.\  {\bf 354} (2001) 243.

\bibitem{adsprelim}
  E.~Ciuffoli, J.~Evslin and F.~Zhao,
  ``Neutrino Physics with Accelerator Driven Subcritical Reactors,''
  JHEP {\bf 1601} (2016) 004
  %doi:10.1007/JHEP01(2016)004
  [arXiv:1509.03494 [hep-ph]].

\bibitem{hondasito}
${\rm{http://www.icrr.u-tokyo.ac.jp/\tilde mhonda/nflx2014/lowe/}}$
 


\bibitem{genie}
  C.~Andreopoulos, A.~Bell, D.~Bhattacharya, F.~Cavanna, J.~Dobson, S.~Dytman, H.~Gallagher and P.~Guzowski {\it et al.},
  ``The GENIE Neutrino Monte Carlo Generator,''
  Nucl.\ Instrum.\ Meth.\ A {\bf 614} (2010) 87
  [arXiv:0905.2517 [hep-ph]].




\bibitem{pregadzooks}
 E.~Kolbe, K.~Langanke and P.~Vogel,
  ``Estimates of weak and electromagnetic nuclear decay signatures for neutrino reactions in Super-Kamiokande,''
  Phys.\ Rev.\ D {\bf 66} (2002) 013007.


\end{thebibliography}
\end{document}